\documentclass[11pt]{article}
\usepackage{jheppub}
\usepackage{graphicx}
\usepackage{amsmath,amssymb,amstext,amsfonts} 
\usepackage{mathrsfs}
\usepackage[utf8]{inputenc}
\usepackage[a4paper,top=1.9in, bottom=0in, left=2in, right=0in]{geometry}
\usepackage{hyphenat}
\usepackage{xcolor}
\usepackage{braket}
\usepackage{array}
\usepackage[bottom]{footmisc}
\usepackage{tikz}
\usepackage{tabularx}
\usepackage{setspace}
\usepackage{enumitem}
\usepackage[numbers]{natbib}
\pdfoutput=1
\usepackage{empheq}
\usepackage{float}
\usepackage{indentfirst}
\usepackage{caption}
\usepackage{subcaption}

\begin{document}

\title{Exotic Black Hole Thermodynamics in Third-Order Lovelock Gravity}

\author[1,2]{Brayden Hull}
\author[3]{Fil Simovic}
\affiliation[1]{Department of Applied Mathematics, University of Waterloo,\\
	Waterloo, Ontario N2L 3G1, Canada}
\affiliation[2]{Perimeter Institute for Theoretical Physics,\\
	31 Caroline St. N.,Waterloo, Ontario N2L 2Y5, Canada}
\affiliation[3]{School of Mathematical and Physical Sciences, Macquarie University,\\
	Sydney, NSW 2109, Australia}
\emailAdd{b2hull@uwaterloo.ca}
\emailAdd{fil.simovic@mq.edu.au}

\abstract{The generalization of Birkhoff's theorem to higher dimensions in Lovelock gravity allows for black hole solutions with horizon geometries of non-constant curvature. We investigate thermodynamic aspects of these `exotic' black hole solutions, with a particular emphasis on their phase transitions. We consider an extended phase space where the cosmological constant acts as a thermodynamic pressure, and examine both uncharged and $U(1)$ charged solutions. In $d=7$, black hole solutions are restricted to having constant curvature horizon base manifolds. Uncharged $d=7$ black holes possess novel triple point phenomena analogous to those recently uncovered in exotic $d=6$ black holes in Gauss-Bonnet gravity, while their charged counterparts generically undergo small-large black hole phase transitions. In $d=8$, we find that both charged and uncharged black holes exhibit triple point behaviour and small-large black hole transitions. We also show that a wide range of `exotic' horizon geometries can be ruled out due to the appearance of naked singularities.}


\maketitle

\vspace{-0.5cm}
\setlength{\parskip}{8pt}

\section{Introduction}

Reconciling the distinctly quantum mechanical nature of Standard Model field theory with the classical notion of gravity as being emergent from the geometry of semi-Riemannian manifolds remains one of the most important open problems in theoretical physics, and while decades of active research have yielded significant progress, much remains to be understood \cite{carlip2015}. One of the most promising paths towards understanding the would-be quantum features of the gravitational interaction has been the study of its thermodynamic properties. The extent to which a thermodynamic description of the gravitational field can be formulated (and understood to arise from its microscopic degrees of freedom) has been explored extensively over the years, both borrowing from and contributing to lessons learned in string theory, quantum field theory in curved space, and information theory. The genesis of the subject can be traced back to the early 1970's when Bardeen, Carter, and Hawking, along with observations by Bekenstein, showed that equilibrium black hole configurations in asymptotically flat spacetime obey a law closely resembling the first law of thermodynamics of ordinary systems \cite{bardeen1973a,bekenstein1973a}. The fact that black holes (which are classically understood to be purely geometric objects) obey such a law has guided research into the thermodynamic properties of black holes ever since \cite{wald2000,carlip2014}.

Somewhat dramatically, black holes have been shown to undergo phase transitions akin to those seen in ordinary thermodynamic systems. The prototypical example is the Hawking-Page transition between an anti-de Sitter black hole and thermal radiation \cite{hawking1983a}. Distinct from the evaporation process, this transition has an important interpretation as a deconfinement transition in gauge theory degrees of freedom through the anti-de Sitter/conformal field theory correspondence, and has provided significant insight into the structure of strongly coupled field theories, the QCD phase diagram, and even unitarity as it relates to the black hole information paradox \cite{maldacena1998,cui2021,harlow2016}. The likeness between black hole thermodynamics and that of traditional matter systems becomes even more salient in the extended phase space (where the cosmological constant is treated as a thermodynamic pressure---see \cite{kastor2009}) where not only do more exotic transitions appear, but also the notion of holographic heat engines and many other interesting phenomena that have been extensively explored over the past few years (see \cite{kubiznak2011} for a review). 

The richest phenomenology is observed in asymptotically anti-de Sitter (AdS) black hole solutions in Lovelock gravity, a higher-dimensional generalization of Einstein gravity which is ordinarily constructed in $d>4$\footnote{Recent attempts at formulating Gauss-Bonnet gravity in $d=4$ indicate it may be possible to obtain nontrivial deviations from general relativity even in this case \cite{glavan2020,hennigar2020a}.} \cite{lovelock1971}. Such black holes have been shown to exhibit not only the ordinary Hawking-Page transition, but also reentrant phase transitions, triple points, superfluid-like transitions, and more \cite{kubiznak2012,altamirano2013,frassino2014,hennigar2017f}. The theory defined by the Lovelock Lagrangian is the most general metric theory of gravity with second-order equations of motion (being free of ghost instabilities), and differs from general relativity through the inclusion of higher-curvature terms to the Einstein-Hilbert action. Lovelock gravity has enjoyed significant theoretical interest, as these corrections to Einstein gravity arise naturally in quantum gravity---for example, the Gauss-Bonnet term appears as the leading correction in 10d gauged supergravity and as a quantum correction from heterotic string theory \cite{cecotti1988}. 

Lovelock gravity possesses another interesting feature that remains largely unexplored in the context of black hole thermodynamics: the evasion of Birkhoff's theorem for general relativity. Ordinarily the well-known Schwarzschild solution 
\begin{equation}
d s^{2}=-f(r, t) d t^{2}+\frac{d r^{2}}{f(r, t)}+r^{2} d \Sigma_{d-2}^{2}
\end{equation}
is unique---although the base manifold $d \Sigma_{d-2}^{2}$ is in principle an arbitrarily chosen Einstein manifold, the $S^2$ is the only possibility in four dimensions. This theorem has a natural generalization to higher dimensions in terms of the Schwarzschild-Tangherlini solution \cite{tangherlini1963}. Birkhoff's theorem also holds in Lovelock gravity \cite{deser2005,zegers2005} where unlike $d=4$ general relativity, there is enough freedom to have arbitrary base manifolds with both constant and non-constant curvature\footnote{In fact, this is a generic feature of higher-dimensional metric theories of gravity whose Lagrangians consist of invariants constructed from the Riemann tensor and its covariant derivatives \cite{MasatoSeiju,hervik2020}.} \cite{dotti2005,ray2015,Dotti_2007,DOTTI_2009,Dotti_2010,Oliva_2013,sourya2015}. In this setting, the constants characterizing the Euler invariants and Lovelock tensors appear in the metric function, and determine the transverse part of the metric. A large new class of black hole solutions therefore exist in Lovelock gravity with `warped' base manifold geometries. The implications of such modifications for the thermodynamic features of black hole solutions are relatively unknown, with the case of Gauss-Bonnet gravity being only very recently examined \cite{hull2021}. In this work, we aim to extend this analysis to third-order Lovelock gravity, where there is significantly more freedom in the choice of horizon geometry for these exotic black holes.

This paper is organized as follows. In Section \ref{review}, we discuss the formulation of Lovelock gravity, and introduce the `exotic' black hole solutions we will study herein. In Section \ref{thermo}, we define the relevant thermodynamic quantities associated with general black hole solutions in Lovelock gravity. In Section \ref{thermo2}, we introduce the solutions considered throughout and derive the relevant thermodynamic quantities. In Section \ref{7d}, we study third-order Lovelock black holes in $d=7$, focusing on both the charged and uncharged solutions. We examine the phase structure through the free energy, and study the equation of state. In Section \ref{8d}, we turn our attention to $d=8$ solutions, the lowest dimension where non-constant curvature base manifolds are allowed. We again examine the phase structure and equation of state, and detail the impact that deviations from constant curvature have on the thermodynamic properties of the solutions. We close in Section \ref{conclusion} with a summary of the main findings and avenues for future investigations. Except for Sections \ref{review} and \ref{thermo}, we use natural units where $c=\hbar=G_N=1$.

\section{Lovelock Gravity and Exotic Black Holes}\label{review}
In this section, we briefly review Lovelock theory and discuss how black hole solutions with exotic horizon geometries arise. 

\subsection{Lovelock Gravity}

Lovelock gravity represents a large class of classical metric gravitational theories that all possess a number of desirable properties, namely, having Lagrangians that are manifestly diffeomorphism-invariant and lead to second-order equations of motion\footnote{This means in particular that there are no linear instabilities in any momentum degrees of freedom in the corresponding Hamiltonian (the Ostragradsky instability).}. Such theories differ from general relativity (which is the unique Lovelock theory in four dimensions\footnote{Attempts to formulate Lovelock gravity in $d=4$ have been made to various degrees of success, see e.g. \cite{glavan2020,hennigar2020a}.}) through the inclusion of higher-curvature terms in the Einstein-Hilbert action, which ordinarily contribute to the equations of motion only in higher dimensions (when $d>4$). The Lovelock Lagrangian is
\begin{equation}\label{lovelockL}
    \mathcal{L}=\frac{1}{16 \pi G} \sum_{k=0}^{K} \hat{\alpha}_{k} \mathcal{L}^{(k)}\ ,
\end{equation}
where $\hat{\alpha}_{k}$ are the Lovelock coupling constants and $\mathcal{L}^{(k)}$ are the Euler densities, which are constructed through various contractions of the Riemann tensor $R_{abcd}$ as 
\begin{equation}
    \mathcal{L}^{(k)}=\frac{1}{2^{k}}\, \delta_{c_{1} d_{1} \ldots c_{k} d_{k}}^{a_{1} b_{1} \ldots a_{k} b_{k}} \,{R_{a_{1} b_{1}}}^{\!c_{1} d_{1}} \ldots {R_{a_{k} b_{k}}}^{\!c_{k} d_{k}}\ .
\end{equation}
With $\mathcal{L}^{(0)}=1$, $\hat{\alpha}_0=-2\Lambda$, $\mathcal{L}^{(1)}=R$ and $\hat{a}_1=1$, the $K=1$ truncation of \eqref{lovelockL} yields the Einstein-Hilbert action. Notably, the $k$-th Euler density is a topological invariant when $d\leq2k$, and therefore does not contribute to the equations of motion of the resulting theory. Nontrivial deviations from general relativity therefore occur only when the dimension of spacetime is large enough for a given finite truncation of the sum \eqref{lovelockL}, namely when $d>2K$.

The full $d$-dimensional action of Lovelock theory with matter can thus be written as
\begin{equation}
    S=\int d^{d} x \sqrt{-g}\left(\frac{1}{16 \pi G_{N}} \sum_{k=0}^{K} \hat{\alpha}_{k} \mathcal{L}^{(k)}+\mathcal{L}_\text{matter}\right)\ ,
\end{equation}
whose variation with respect to the metric yields the equations of motion
\begin{equation}
    \sum_{k}^{K} \hat{\alpha}_{k} \mathcal{G}_{a b}^{(k)}=8 \pi G_{N} T_{a b}\ ,
\end{equation}
where $T_{a b}$ is the stress-energy tensor and $\mathcal{G}^{(k)}_{ab}$ are the Lovelock tensors, given by
\begin{equation}
    \mathcal{G}_{ ab}^{(k)}=-\frac{1}{2^{(k+1)}}\, g_{z a}\, \delta_{b e_{1} f_{1} \ldots e_{k} f_{k}}^{z c_{1} d_{1} \ldots c_{k} d_{k}}\, {R_{c_{1} d_{1}}}^{e_{1} f_{1}} \ldots {R_{c_{k} d_{k}}}^{e_{k} f_{k}}\ .
\end{equation}
We will be interested in both uncharged and $U(1)$ charged black holes, the latter involving an additional electromagnetic term in the action: $\mathcal{L}_{\text{matter}} = -4 \pi G_{N} F_{a b} F^{a b}$. The resulting field equations are then
\begin{equation} \label{fieldeq}
\sum_{k=0}^{K} \hat{\alpha}_{(k)} \mathcal{G}_{a b}^{(k)}=8 \pi G_{N}\left(F_{a c} F_{b}^{c}-\frac{1}{4} g_{a b} F_{c d} F^{c d}\right).
\end{equation}
We will use the following spherically symmetric ansatz for the metric and gauge field:
\begin{equation}\label{metric}
      ds^2 = 
{\textsf{g}}_{ij} {dy^i dy^j} 
    +  {\gamma}_{\alpha\beta}dx^\alpha dx^\beta
    =-f(r)dt^2 + \frac{dr^2}{f(r)} + r^2 d\Sigma_{d-2}^2 \ ,
    \quad
    F = \frac{Q}{r^{d-2}}dt\wedge dr\ . 
\end{equation}
Latin indices are reserved for coordinates in the $t-r$ plane while Greek indices refer to coordinates on the base manifold $d\Sigma_{d-2}^2$. The importance of this metric is how one defines $\gamma_{\alpha \beta}$, which determines the geometry of the transverse sections (including the horizon). In this work we take the base manifold with metric $\gamma_{\alpha \beta}$ to be a $(d-2)$ compact space not necessarily of constant curvature. The constant curvature case (corresponding to `ordinary' Lovelock black holes) has been studied extensively \cite{frassino2014}, while the non-constant curvature case has only been examined in the context of Gauss-Bonnet gravity \cite{hull2021}. Inserting the metric \eqref{metric} into the field equations produces two simplified expressions:
\begin{equation}
    \begin{split}
    \mathcal{G}_{j}^{i} & \equiv \frac{-(d-2) \delta_{j}^{i}}{2 r^{d-2}} \frac{d}{d r} \sum_{n=0}^{K}\left\{b_{n}\left(r^{d-2 n-1} A_{n}\left(\frac{-f(r)}{r^{2}}\right)\right)\right\}=8 \pi G_{N} T_{j}^{i}\ ,\\
    \mathcal{G}^{\alpha}_{\beta} & \equiv \frac{- \delta^{\alpha}_{\beta}}{2 r^{d-3}} \frac{d^2}{dr^2} \sum_{n=0}^{K} \left\{ b_{n}\left( r^{d-2 n-1} A_{n}\left(\frac{-f(r)}{r^{2}}\right) \right) \right\} = 8 \pi G_{N} T^{\alpha}_{\beta}\ ,
    \end{split} \label{fieldequations}
\end{equation}
where
\begin{equation}
    A_{n}\left(\frac{-f(r)}{r^{2}}\right) \equiv \sum_{k=n}^{K} \alpha_{k} \binom{k}{n} \left( \frac{-f(r)}{r^{2}}\right)^{k-n} \ ,
\end{equation}
and the $b_n$ are arbitrary constants which depend on the geometry of the base manifold $d\Sigma_{d-2}^2$ (see Section \ref{exotichor}). These field equations determine the metric function $f(r)$, which is obtained as a solution to the following polynomial of degree $K$ in $f(r)$:
\begin{equation}\label{poly}
    \sum_{n=0}^{K} \frac{b_{n}}{r^{2n}} \left( \sum_{k=n}^{K} \alpha_{k} \binom{k}{n}\left(\frac{-f(r)}{r^{2}}\right)^{k-n}\right )=\frac{16 \pi G_{N} M}{(d-2) \Sigma_{d-2} r^{d-1}}-\frac{8 \pi G_{N} Q^{2}}{(d-2)(d-3)} \frac{1}{r^{2 d-4}}
\end{equation}
Here $M$ represents the ADM mass, and $Q$ is the charge defined by 
\begin{equation}
    Q=\frac{1}{2 \Sigma_{d-2}}\int \star\, F\ .
\end{equation}
Finally, $\alpha_{k}$ are the re-scaled Lovelock coupling constants defined by
\begin{equation}
\alpha_{0}=\frac{\hat{\alpha}_{(0)}}{(d-1)(d-2)} = 
\quad \alpha_{1}=\hat{\alpha}_{(1)}, \quad \alpha_{k}=\hat{\alpha}_{(k)} \prod_{n=3}^{2 k}(d-n) \quad \text { for } \quad k \geq 2\ .
\end{equation}
Note that constant curvature base manifolds have $b_n=\kappa^n$ with $\kappa\in\{-1,0,+1\}$, corresponding to negative, flat and positive curvature respectively. This can be shown by direct comparison of \eqref{poly} with the familiar form of the constant curvature $f(r)$ polynomial given by
\begin{equation}
    \sum_{k=0}^{K} \alpha_{k}\left(\frac{\kappa-f}{r^{2}}\right)^{k}=\frac{16 \pi G_{N} M}{(d-2) \Sigma_{d-2}^{(\kappa)} r^{d-1}}-\frac{8 \pi G_{N} Q^{2}}{(d-2)(d-3)} \frac{1}{r^{2 d-4}}\ .
\end{equation}
where $M$, $Q$ and the coupling constants are as defined above, and $\Sigma_{d-2}^{(\kappa)}$ is the volume of the constant curvature compact space.

In what follows we take $K=3$ in the above expressions, corresponding to third-order Lovelock gravity, and set $\alpha_{1}=1$ in order to recover general relativity in the $\{\alpha_2,\alpha_3\}\rightarrow 0$ limit. Useful for analyzing the structure of the resulting solutions is the Kretschmann scalar, a curvature invariant which for our general metric \eqref{metric} can be written as
\begin{align}
   \mathcal{K}&= R^{ abcd }R_{abcd }\\
   &=\left(\frac{d^{2} f(r)}{d r^2}\right)^{2} +  \frac{2(d-2)}{r^2}\left( \frac{d f(r)}{dr} \right) ^2 + \frac{2(d-2)(d-3) f(r)^2}{r^4}  - \frac{4 R[\gamma] f(r)}{r^4} +\frac{\mathcal{K}[\gamma]}{r^4}\ .\nonumber
\end{align}
It is a function of both the metric function $f(r)$ and the Ricci and Kretschmann scalars of the base manifold ($R[\gamma]$ and $\mathcal{K}[\gamma]$, respectively), and provides a coordinate-free measure of curvature which can be used to locate singularities and other pathologies in the spacetime.

\subsection{Exotic Horizon Geometries}\label{exotichor}

The primary motivation of this work is to examine the role that horizon geometry plays in determining the thermodynamic features of black hole spacetimes. Ordinarily, Birkhoff's theorem in general relativity strictly limits the types of allowed horizon geometries. For a general spherically symmetric metric of the form
\begin{equation}
d s^{2}=-f(r, t) d t^{2}+\frac{d r^{2}}{f(r, t)}+r^{2} d \Sigma_{d-2}^{2}\ ,
\end{equation}
the transverse part of the metric $d \Sigma_{d-2}^{2}$ must be an Einstein manifold (one for which $R_{\mu\nu}=kg_{\mu\nu}$ in local coordinates for some constant $k$) with spherical symmetry\footnote{Absent spherical symmetry one may also consider $\mathcal{H}^2$, $\text{I\!R}^2$, $T^2$, $\text{I\!R}P^2$, and others.}. In $d=4$ there is only one possibility, namely   $\mathcal{S}^2$, implying that all vacuum solutions with $SO(3)$ symmetry and positive curvature are diffeomorphic to the maximal extension of the Schwarzschild solution in an open subset. In higher dimensions other Einstein manifolds are allowed, such as the Bohm metrics [ref].

In Lovelock gravity the base manifold $d \Sigma_{d-2}^{2}$ need not be an Einstein manifold, allowing for a rich variety of horizon geometries \cite{ray2015}. In fact, $d \Sigma_{d-2}^{2}$ can be rather arbitrary, the only requirement being that various constraints are satisfied which relate the Lovelock tensors of the base manifold. Specifically, the metric function $f(r)$ is a solution to the polynomial \eqref{poly} with the additional requirement that
\begin{equation}\label{lovelockbase}
    \hat{\mathcal{G}}_{\beta}^{(n) \alpha}=-\frac{(d-3) !\, b_{n}}{2(d-2 n-3) !} \delta_{\beta}^{\alpha}\ ,
\end{equation}
where $\hat{\mathcal{G}}_{\beta}^{(n) \alpha}$ are simply the Lovelock tensors of the base manifold. The constants $b_n$ appear in the polynomial \eqref{poly} defining the metric function and are arbitrary, though they depend on the geometry of the base manifold through \eqref{lovelockbase}.

When $d=2K+1$ the tranverse space must be of constant curvature \cite{Dotti_2007,farhangkhah2014}, so the first dimension in which the $K$-th Lovelock term begins to contribute only admits spherical, flat, and hyperbolic transverse sections and has $b_{n}=\kappa^{n}$ with $\kappa\in\{-1,0,+1\}$. For third-order Lovelock gravity, this means that only black hole solutions in $d\geq 8$ can possess exotic horizon geometries. Furthermore, we emphasize that significant restrictions on the allowed geometries arise from the field equations. For example, the Bohm metrics 
\begin{equation}\label{bohm}
    d s^{2}=d \rho^{2}+a(\rho)^{2} d \Omega_{p}^{2}+b(\rho)^{2} d \Omega_{q}^{2}
\end{equation}
represent an infinite family of local metrics that can be extended to manifolds of topology $S^{p+q+1}$. Black holes with Bohm horizons have non-constant curvature, yet of the infinitely many metrics given by different choices of $a(\rho)$ and $b(\rho)$ in \eqref{bohm}, only one choice turns out to be admissible as a horizon due to the Gauss-Bonnet correction appearing when $K\geq 2$ \cite{dotti2005}. Higher orders in Lovelock theory likely further constrain the allowed choices of horizon geometry, though this remains a currently unexplored feature of exotic black holes in higher-dimensional Lovelock gravity. In this work we focus on choices of $b_n$ that produce everywhere real and analytic metric functions, ensuring that curvature scalars are finite and that the spacetime has the appropriate asymptotics, rather than attempting to derive general constraints on the $b_n$s. We will see that many interesting features appear even in the limited parameter space we study herein.

\section{Extended Phase Space Thermodynamics} \label{thermo}

We operate in an {\it extended} thermodynamic phase space, which arises from considering variations of the cosmological constant $\Lambda$ when constructing the first law for black hole systems. The inclusion of such a variation is motivated by the observation that the `ordinary' first law of black hole mechanics contains no pressure-volume term, since there are no obvious notions of such quantities for a black hole. Since the cosmological constant is equivalent to a perfect fluid stress-energy tensor, a variable $\Lambda$ provides a mechanism for would-be pressure variations in a gravitational system. Precisely, we identify the cosmological constant with a fluid pressure through the relation
\begin{equation}\label{pressure}
P=-\frac{\Lambda}{8 \pi }=-\frac{(d-1)(d-2)}{16 \pi  l^{2}}\ ,
\end{equation}
where $l$ is the AdS length scale. The consequences of such a variation were first explored by Kastor and collaborators \cite{kastor2009,kastor2010a,kastor2011b}, who showed through a Hamiltonian analysis that the resulting first law is supplemented by a term $\Theta\delta\Lambda$, where the potential $\Theta$ conjugate to $\Lambda$ plays the role of the thermodynamic volume, and, at least for Schwarzschild-AdS black holes, is exactly the renormalized geometric volume of AdS minus the volume of the black hole. In this way, a pressure-volume term naturally enters into the first law. From this extended phase space emerged the area of research known as black hole chemistry, so named because the inclusion of pressure in the thermodynamic description gives rise to a number of phenomena in the gravitational sector that bear striking resemblance to thermodynamic phenomena observed in everyday systems such as water, helium, and dry ice \cite{hennigar2017f,kubiznak2017b}. The first law in the extended phase space formally reads
\begin{equation}
\delta M=T \delta S+V \delta P\ ,
\end{equation}
where additional terms enter if angular momentum or charge are present. In Lovelock theory, the mass $M$, temperature $T$, entropy $S$, and charge $Q$ of static, asymptotically AdS black holes are given by \cite{hull2021}
\begin{align}
\quad M&=\frac{\Sigma_{d-2}(d-2)}{16 \pi G_{N}} \sum_{k=0}^{K} \alpha_{k} b_{k} r_{+}^{d-1-2 k}+\frac{\Sigma_{d-2}}{2(d-3)} \frac{Q^{2}}{r_{+}^{d-3}}\ , \\
S&=\frac{\Sigma_{d-2}(d-2)}{4 G_{N}} \sum_{k=0}^{K} \frac{k b_{k-1} \alpha_{k} r_{+}^{d-2 k}}{d-2 k}\label{entropy1}\ ,\\
T&=\frac{1}{4 \pi r_{+} D\left(r_{+}\right)}\left[\sum_{k=0}^{K} b_{k} \alpha_{k}(d-2 k-1) r_{+}^{-2(k-1)}-\frac{8 \pi G_{N} Q^{2}}{(d-2) r_{+}^{2(d-3)}}\right]\ , \label{temp}
\end{align}
where $r_+$ is the largest positive real root of $f(r_+)=0$ and we have defined
\begin{equation}
    D\left(r_{+}\right)\equiv\sum_{k=1}^{K} k \alpha_{k} b_{k-1} r_{+}^{-2(k-1)}\ .
\end{equation}
These quantities obey the following first law and Smarr relation:
\begin{equation}
    \begin{split}
    \delta M &=T \delta S-\frac{1}{16 \pi G_{N}} \sum_{h} \hat{\Psi}^{(k)} \delta \hat{\alpha}_{(k)}+\Phi \delta Q\ , \\
    (d-3) M &=(d-2) T S+\sum_{k} 2(k-1) \frac{\hat{\Psi}^{(k)} \hat{\alpha}_{(k)}}{16 \pi G_{N}}+(d-3) \Phi Q\ .
    \end{split}
\end{equation}
Here, $\hat{\Psi}^{(k)}$ are potentials conjugate to the Lovelock coupling constants $\hat{\alpha}_{(k)}$ which are promoted to thermodynamic variables. The potentials are defined by
\begin{equation}
    \hat{\Psi}^{(k)}=4 \pi T \mathcal{A}^{(k)}+\mathcal{B}^{(k)}+\Theta^{(k)}\ ,
\end{equation}
with 
\begin{equation}
\begin{aligned}
\mathcal{B}^{(k)} &=-\frac{16 \pi k G_{N} M(d-1) !}{b(d-2 k-1) !}\left(-\frac{1}{\ell^{2}}\right)^{k-1}, \quad b=\sum_{k} \frac{\hat{\alpha}_{k} k(d-1) !}{(d-2 k-1) !}\left(-\frac{1}{\ell^{2}}\right)^{k-1}\ , \\
\Theta^{(k)} &=\int_{\Sigma} \sqrt{-g} \mathcal{L}^{(k)}[s]-\int_{\Sigma_{\mathrm{AdS}}} \sqrt{-g_{\mathrm{AdS}}} \mathcal{L}^{(k)}\left[s_{\mathrm{AdS}}\right]\ .
\end{aligned}
\end{equation}
The thermodynamic entropy is also modified in a non-trivial way compared to the Einstein-AdS case, and is no longer one-quarter the area of the horizon. It is now given by
\begin{equation}\label{entropy}
    S=\frac{1}{4 G_{N}} \sum_{k} \hat{\alpha}_{k} \mathcal{A}^{(k)}, \quad \mathcal{A}^{(k)}=k \int_{\mathcal{H}} \sqrt{\sigma} \mathcal{L}^{(k-1)}\ ,
\end{equation}
where $\sigma$ is the induced metric on the black holes horizon and $\mathcal{L}^{(k-1)}$ are the Euler densities of the horizon.

\section{Exotic Black Holes in Third-Order Lovelock Gravity} \label{thermo2}
In this section, we present general solutions for charged black hole metrics in third-order Lovelock gravity with exotic horizon geometries. We solve the polynomial \eqref{poly} to determine the metric function, and examine various constraints on the couplings $\{\alpha_i\}$ that arise from considerations of analyticity and asymptotic behaviour. We also determine the equation of state and free energy, and place bounds on the allowed pressure (cosmological constant) for these black holes. 

\subsection{Solutions}
Setting $K=3$ in \eqref{poly} gives the following polynomial for the metric function $f$:
\begin{align}\label{fpoly}
    -\frac{\alpha_{3} f^{3}}{r^{6}}+\left(\frac{\alpha_{2}}{r^{4}}+\frac{3 b_{1} \alpha_{3}}{r^{6}}\right) f^{2}&+\left(-\frac{1}{r^{2}}-\frac{2 b_{1} \alpha_{2}}{r^{4}}- \frac{3 b_{2} \alpha_{3}}{r^{6}}\right) f +\alpha_{0}+\frac{b_{1}}{r^{2}}+& \frac{b_{2} \alpha_{2}}{r^{4}}+\frac{b_{3} \alpha_{3}}{r^{6}}=A(r)\ ,\nonumber
    \\ A(r) &\equiv \frac{16 \pi  M }{(d-2) \Sigma_{d-2} r^{d-1}}-\frac{8 \pi  Q^{2} }{(d-2)(d-3)r^{2d-4}}\ .
\end{align}
This is a cubic equation for $f$ possessing three distinct solutions. The first is given by
\begin{equation}\label{ebranch}
f_E=\frac {  \Big( r^4\left(3 \alpha_3+\alpha_2^2\right)- 9\left(b_1^2-b_2\right) {\alpha_3}^{2} \Big)+\left( r^2\alpha_2+3 \alpha_3 b_1\right) \left(\tfrac{1}{2}\big(Y+\sqrt {X}\big)\right)^{1/3}-\left( \tfrac{1}{2}\big(Y+\sqrt {X}\big) \right) ^{2/3}}{3 \alpha_3\left({\tfrac{1}{4}\big(Y+\sqrt {X}}\big)\right)^{1/3}}
\end{equation}
where we have defined
\begin{align*}
X\equiv&\ \Big(27 \alpha_3^3 \big(2 b_1^3-3 b_1 b_2+b_3\big)-\big(27\alpha_3^2\big(A(r)-\alpha_0\big)-2 \alpha_2^3 +9 \alpha_2 \alpha_3  \big) r^6\Big)^2\nonumber \\
&\ +4 \Big(9 \alpha_3^2 \big(b_2-b_1^2\big)-\big(\alpha_2^2 +3 \alpha_3 \big)r^4\Big)^3\ ,\nonumber \\
Y\equiv&\ (81 b_1 b_2-54 b_1^3-27 b_3) \alpha_3^3+\Big(27\alpha_3^2 \big(A(r)-\alpha_0\big)+9 \alpha_2 \alpha_3 -2 \alpha_2^3\Big) r^6\nonumber\ .
\end{align*}
The subscript $E$ indicates that this branch gives the appropriate Einstein limit when successively taking the limits $\alpha_3\rightarrow0$ and $\alpha_2\rightarrow 0$. The other two (conjugate) branches are given by the solutions
\begin{footnotesize}\begin{equation}\label{otherbranches}
    f=\frac {\left(1\mp i\sqrt {3}
	\right) \Big( r^4\!\left(3 \alpha_3+\alpha_2^2\right)- 9\left( b_1^2-b_2 \right) \alpha_3^{2} \Big)+\left( r^2\alpha_2+3\alpha_3 b_1\right)\! \sqrt[3]{4\big(Y+\sqrt {X}\big)}+ \left(1\pm i\sqrt {3} \right) \left( \tfrac{1}{2}\big(Y+\sqrt {X}\big) \right) ^{2/3} }{3 \alpha_3\sqrt[3]{4\big(Y+\sqrt {X}}\big)}\nonumber.
\end{equation}\end{footnotesize}
\vspace{-7pt}

These two solutions are distinguished by a lack of convergent $\alpha_2\rightarrow0$ limit. We further differentiate between the two by considering their Gauss-Bonnet ($\alpha_3\rightarrow0$) limits. The solution $f_{GB}$, given by the upper sign, converges in this limit and yields the appropriate metric function for the Schwarzschild-AdS black hole (when $\{b_1,b_2,b_3\}=\{\kappa,\kappa^2,\kappa^3\}$) and is therefore referred to as the Gauss-Bonnet branch. The solution $f_L$, given by the lower sign, has a divergent $\alpha_3\rightarrow0$ limit and exists only in third-order Lovelock gravity. We refer to this as the Lovelock branch.

\begin{figure}[H]
    \includegraphics[width = 0.5\textwidth]{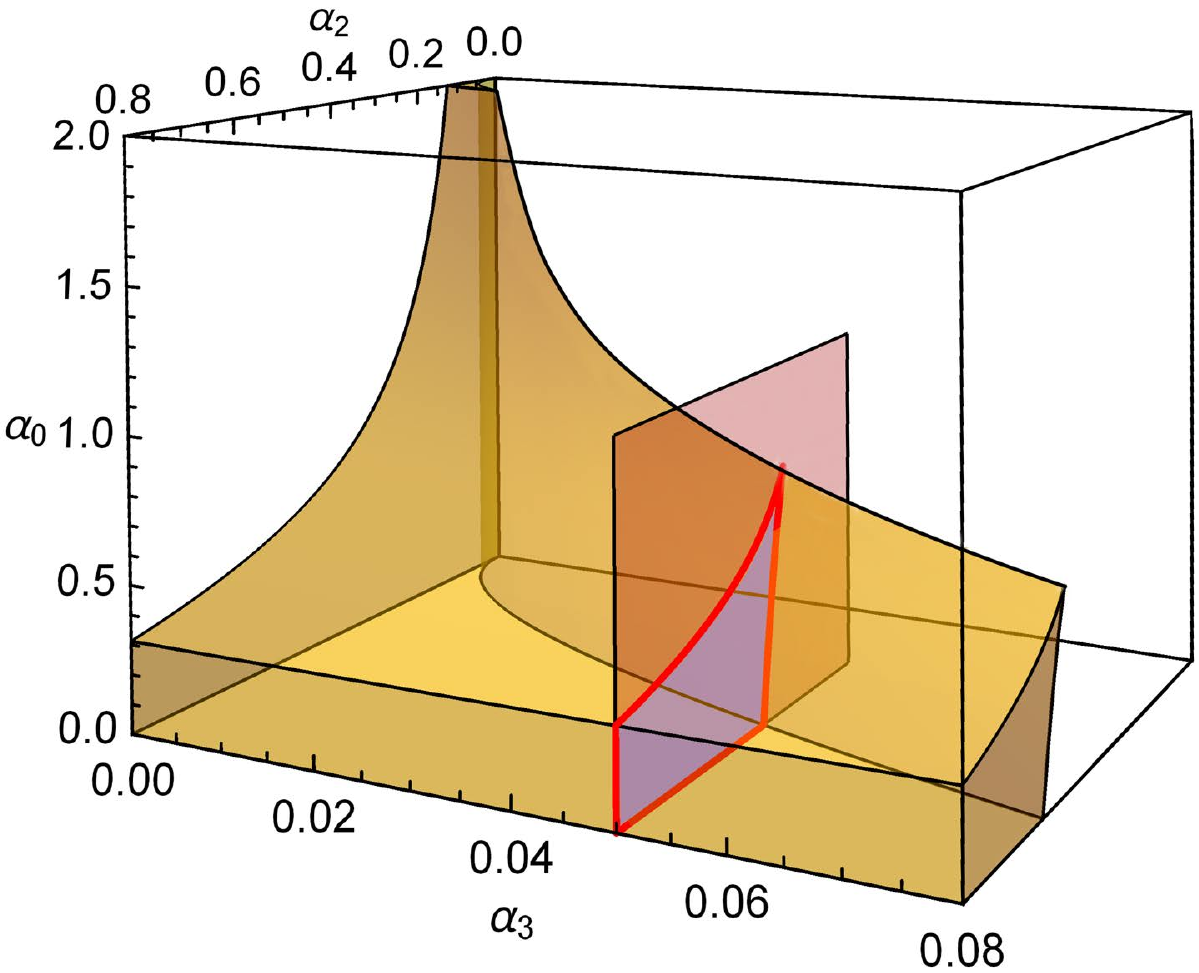}\qquad\includegraphics[width = 0.45\textwidth]{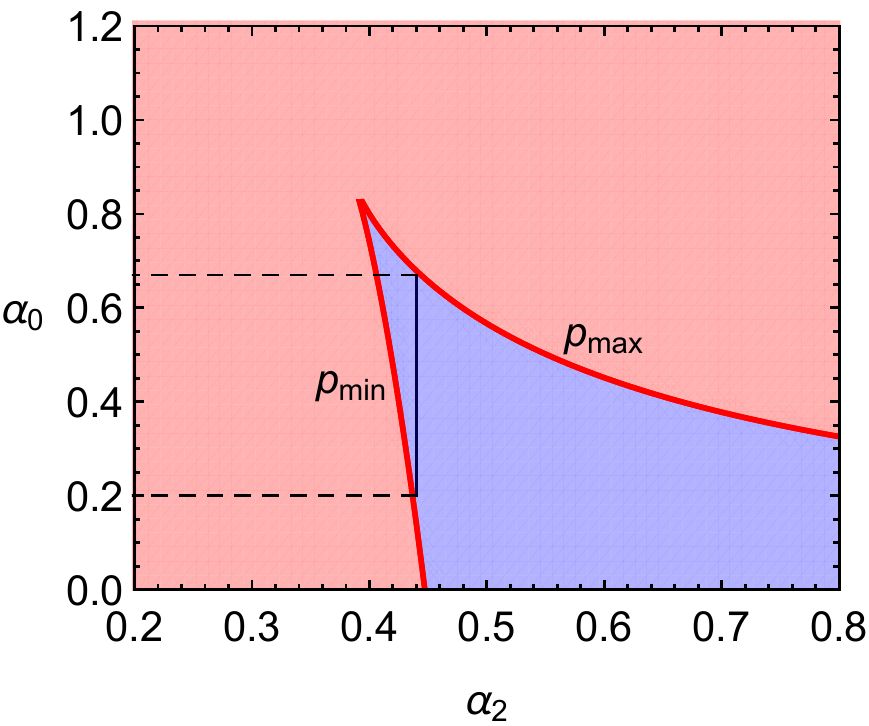}
    \caption{The discriminant condition for solutions to \eqref{fpoly}. \textbf{Left:} The shaded area indicates regions of the parameter space where all solutions are real, as determined by the constraint \eqref{cond1}.  \textbf{Right:} A fixed-$\alpha_3$ slice taken from the plot on the left, showing the emergence of a minimum/maximum allowed pressure for certain values of $\alpha_2$.}
    \label{fig:alpharegion}
\end{figure}

A number of considerations arise from the form of the solutions \eqref{ebranch}--\eqref{otherbranches}. As discussed in \cite{frassino2014}, a bound on the couplings $\{\alpha_2,\alpha_3\}$ can be obtained by demanding that all three solutions to \eqref{fpoly} are simultaneously real-valued. This is determined by positivity of the discriminant of the cubic equation for $f$, which implies the following constraint on the couplings:
\begin{equation}\label{cond1}
    27 {\alpha_{0}}^{2} {\alpha_{3}}^{2}+4 \alpha_{0} \alpha_{2}^{3}-18 \alpha_{0} \alpha_{2} \alpha_{3}-{\alpha_{2}}^{2}+4 \alpha_{3} \leq 0
\end{equation}
When \eqref{cond1} is satisfied, one can smoothly take the Einstein limit by appropriate choice of solution. In \figurename{ \ref{fig:alpharegion}}, we plot regions in the parameter space for which this condition is satisfied.

The inequality \eqref{cond1} puts an upper (and sometimes lower) bound on the allowed pressure for certain values of $\alpha_2$, given in the Einstein and Gauss-Bonnet cases by
\begin{equation}
    P_{\pm}=\frac{(d-1)(d-2)}{432 \pi }\frac{9 \alpha_{2} \alpha_{3} - 2\alpha_{2}^{3} \pm 2 \left( \alpha_{2}^{2}-3\alpha_{3}\right)^\frac{3}{2} }{\alpha_{3}^2}\ .
\end{equation}
Now introducing the following dimensionless variables
\begin{equation} 
    \alpha=\frac{\alpha_{2}}{\sqrt{\alpha_{3}}}\ , \quad p=4 \sqrt{\alpha_{3}} P\ ,r_{+}=v \alpha_{3}^{\frac{1}{4}}, \quad T=\frac{t \alpha_{3}^{-\frac{1}{4}}}{d-2}, \quad m=\frac{16 \pi M}{(d-2) \Sigma_{d-2}^{(\kappa)} \alpha_{3}^{\frac{d-3}{4}}}, \quad Q=\frac{q}{\sqrt{2}} \alpha_{3}^{\frac{d-3}{4}}
\end{equation}
the pressure can be expressed as
\begin{equation}\label{pminmax}
    p_{\pm}= \frac{(d-1)(d-2)}{108 \pi} \left( 9\alpha - 2 \alpha^{3} \pm 2\left(\alpha^{2} - 3 \right)^{\frac{3}{2}}\right)\ .
\end{equation}
This matches our expectations for the constant curvature case, since these bounds are sensitive only to the asymptotic structure of the solutions that are independent of the horizon geometry. We see that both branches terminate at the same values of $\alpha=\sqrt{3}$, below which the pressure becomes imaginary and the spacetime is no longer asymptotically anti-de Sitter. We therefore limit our analysis to couplings for which $\alpha \geq \sqrt{3}$. There is also a narrow region in the parameter space where a {\it minimum} pressure exists along with a maximum; in this case when $\sqrt{3}<\alpha<2$. When $\alpha\geq 2$ there is no lower bound on the pressure.

\subsection{Equation of State and Free Energy}
For ordinary thermodynamic systems, the equation of state provides a closure relation between thermodynamic degrees of freedom, and is usually phenomenological in nature. For black holes, the equation of state can be constructed directly from \eqref{pressure} and \eqref{temp}. For $K=3$, we have that
\begin{align}
    P=&\ \frac{(d-2) T}{4 r_+}-\frac{(d-2) (d-3)b_{1}}{16 \pi r_+^{2}}+\frac{(d-2) \alpha_{2} b_{1} T}{2 r_+^{3}} -\frac{(d-2) (d-5)\alpha_{2} b_{2}}{16 \pi r_+^{4}}\nonumber \\
    &+ \frac{3(d-2) \alpha_{3} b_{2} T}{4 r_+^{5}}-\frac{(d-2)(d-7)\alpha_{3} b_{3}}{16 \pi r_+^{6}}+\frac{Q^{2}}{2 r_+^{2 d -4}}\ ,
\end{align}
and in terms of the dimensionless variables defined above we have
\begin{align}\label{eos1}
      p=&\ \frac{t}{v}-\frac{(d-2)(d-3) b_{1}}{4 \pi v^{2}}+\frac{2 \alpha b_{1} t}{ v^{3}}-\frac{(d-2)(d-5) \alpha b_{2}}{4  \pi v^{4}}+\frac{3 b_{2} t}{v^{5}} \nonumber \\
      & - \frac{(d-2)(d-7)b_{3}}{4 \pi v^{6}} + \frac{q^{2}}{v^{2d-4}}\ .
\end{align}
From the equation of state, critical points are easily identified as locations where
\begin{equation}\label{criticalpoints}
    \frac{\partial p}{\partial v}=0\ , \quad \frac{\partial^2 p}{\partial v^2}=0\ .
\end{equation}
The first of these conditions yields the critical temperature $t_{c}$:
\begin{equation}
    t_{c}=\frac{(d-2)\left(-4 v_{c}^{10-2 d} q^{2} \pi+b_{1}(d-3) v_{c}^{4}+2 \alpha b_{2}(d-5) v_{c}^{2}+3b_{3} (d-7)\right)}{2 \pi\left(6 \alpha b_{1} v_{c}^{2}+v_{c}^{4}+15 b_{2}\right) v_{c}}
\end{equation}
The second condition determines the critical volume $v_{c}$ as a solution to the following polynomial:
\begin{align}\label{vc}
        0=&\ 4 \pi v_{c}^{10-2 d}\left[(2 d-5) v_{c}^{4}+6\alpha b_{1}(2 d-7) v_{c}^{2}+15 b_{2}(2 d -9)\right](d-2) q^{2}\\
        & + (d-2) \left[ 6\alpha ((d-3)b_{1}^{2}-b_{2}(d-5))v_{c}^{6} - b_{1}(d-3)v_{c}^{8} - 45 b_{3} b_{2}(d-7)\right] \nonumber\\
        & + (d-2) \left[6 \alpha\left(5 b_{2}^{2}(d-5) -9(d-7) b_{1} b_{3} \right) v_{c}^{2} - 3\left( b_{2}\left( 4(d-5)\alpha^{2} - 15 d + 45 \right)b_1\right.\right.\nonumber\\
        & + \left.\left.5 b_{3}(d -7) \right)v_{c}^{4}\right]\nonumber
\end{align}
Once the critical volume and temperature have been found, the equation of state \eqref{eos1} determines the critical pressure $p_{c}$. As \eqref{vc} is generally a high-degree polynomial in $v_c$, multiple solutions may exist. We restrict our analysis to choices of parameters for which $\{v_{c},t_{c},p_{c}\}$ are all positive and real. Additional restrictions on the parameter space arise from entropic considerations, namely the positivity of \eqref{entropy} which implies that
\begin{equation}\label{entropybound}
  (d-6)(d-4) v^{4}+2  b_{1} \alpha(d-2)(d-6)v^{2}+ 3 (d-2)(d-4) b_{2} > 0\ .
\end{equation}
Since $d \geq 7$ and $\alpha>0$, this is automatically satisfied when $b_1\geq 0$ and $b_2\geq 0$.

The Gibbs free energy $G$ and its dimensionless counterpart $g$ can likewise be constructed. The equilibrium state of the system globally minimizes the Gibbs free energy, and coexistence phases are easily identified as locations where $r_+(G)$ is multivalued. Defining the dimensionless Gibbs free energy as
\begin{equation}
    g=\frac{1}{\Sigma_{d-2}} \alpha_{3}^{\frac{3-d}{4}} G\ ,
\end{equation}
we have 
\begingroup
\addtolength{\jot}{0.7em}
\begin{align}
     g  =&\ \frac{q^{2}\!\left((d-4)\left(\left(v^{4}+3 b_{2}\right) d^{2}-\left(\frac{17 v^{4}}{2}+\frac{39 b_{2}}{2}\right)\! d+15 v^{4}+27 b_{2}\right)+2 b_{1}\!\left(d-\frac{7}{2}\right)(d-2) \alpha\, v^{2}(d-6)\right) }{2(d-3)(d-2)(d-4)\left(2 \alpha v^{2} b_{1}+v^{4}+3 b_{2}\right)(d-6)v^{d-3}} \nonumber\\
     & -\frac{p\left(15 b_{2}(d-2)(d-4) v^{d-1}+(d-6)\left(6 b_{1} \alpha(d-2) v^{d+1}+v^{d+3}(d-4)\right)\right)}{4(d-6)(d-2)(d-4)\left(2 \alpha v^{2} b_{1}+v^{4}+3 b_{2}\right)(d-1)}\nonumber \\ 
     & + \frac{3 b_{2} b_{3}(d-2)(d-4) v^{d-6}+6\left(b_{3}(d-6) b_{1}-\frac{b_{2}^{2}(d-4)}{2}\right) \alpha(d-2) v^{d-4}}{16\pi(d-6) v(d-4) \left(2 \alpha v^{2} b_{1}+v^{4}+3 b_{2}\right)} \nonumber\\
     & + \frac{\left((d-2) b_{2}\left(36-9 d\right) b_{1}+5 b_{3}(d-4)(d-6)+2(d-2) b_{2}\left(\alpha^{2} d-6 \alpha^{2}\right) b_{1}\right) v^{d-2}}{16\pi(d-6) v(d-4) \left(2 \alpha v^{2} b_{1}+v^{4}+3 b_{2}\right)}\nonumber \\
     & - \frac{ \alpha\left(2(d-2) b_{1}^{2}-3\, b_{2}(d-4)\right) v^{d}-b_{1}(d-4) v^{d+2}}{16\pi v(d-4) \left(2 \alpha v^{2} b_{1}+v^{4}+3 b_{2}\right)}\ .
\end{align}
\endgroup
\\
This can be plotted parametrically against the equilibrium temperature of the system, given by 
\begin{equation}
    t=\frac{-4 v^{10-2 d} q^{2} \pi+4 v^{6} p \pi+b_{1}(d-2)(d-3) v^{4}+\alpha b_{2}(d-2)(d-5) v^{2}+b_{3}(d-2)(d-7)}{4 \pi\left(2 \alpha v^{2} b_{1}+v^{4}+3 b_{2}\right) v}
    \end{equation}
and using the horizon radius $r_+$ as the parameter. The black hole size thus serves as the order parameter characterizing distinct equilibrium states and/or phases of the system. One can verify that these quantities agree with known results for spherical and hyperbolic black holes with constant curvature horizons \cite{frassino2014}. Our analysis will be predicated on the free energy and equation of state, which provide complementary information about the location and nature of critical points of the black hole system. We will see how moving past the restriction of constant curvature transverse spaces markedly alters the nature of the transitions present in these systems.

\section{Thermodynamics of $d=7$ Exotic Black Holes}\label{7d}

In the following sections we examine thermodynamic aspects of the general solutions presented above, focusing on seven and eight-dimensional charged and uncharged examples. We study the equation of state and thermodynamic free energy in each case, and locate critical points which separate different phases of the system. We work in the canonical ensemble where the physical charge is fixed. Seven-dimensional black holes in third-order Lovelock gravity do not admit horizons of non-constant curvature, so this case corresponds to `ordinary' black holes that have been studied previously \cite{frassino2014}. Nonetheless, we reexamine this case and discover new features previously unseen for this class of black holes. 
\\

In seven dimensions, we are restricted to constant curvature horizon cross sections. This arises from a dimensional analysis of the field equations and the requirement that the base manifold be of constant curvature \cite{farhangkhah2014}. This restricts our topological terms to be $b_{1}=b_{3}=\pm 1$ and $b_{2}=1$ (we may also have $b_{n}=0\  \forall n$ but we will not be examining this case). We consider the charged and uncharged cases separately.

\subsection{Uncharged Solutions}
\begin{figure}[H]
    \includegraphics[width=0.325\textwidth]{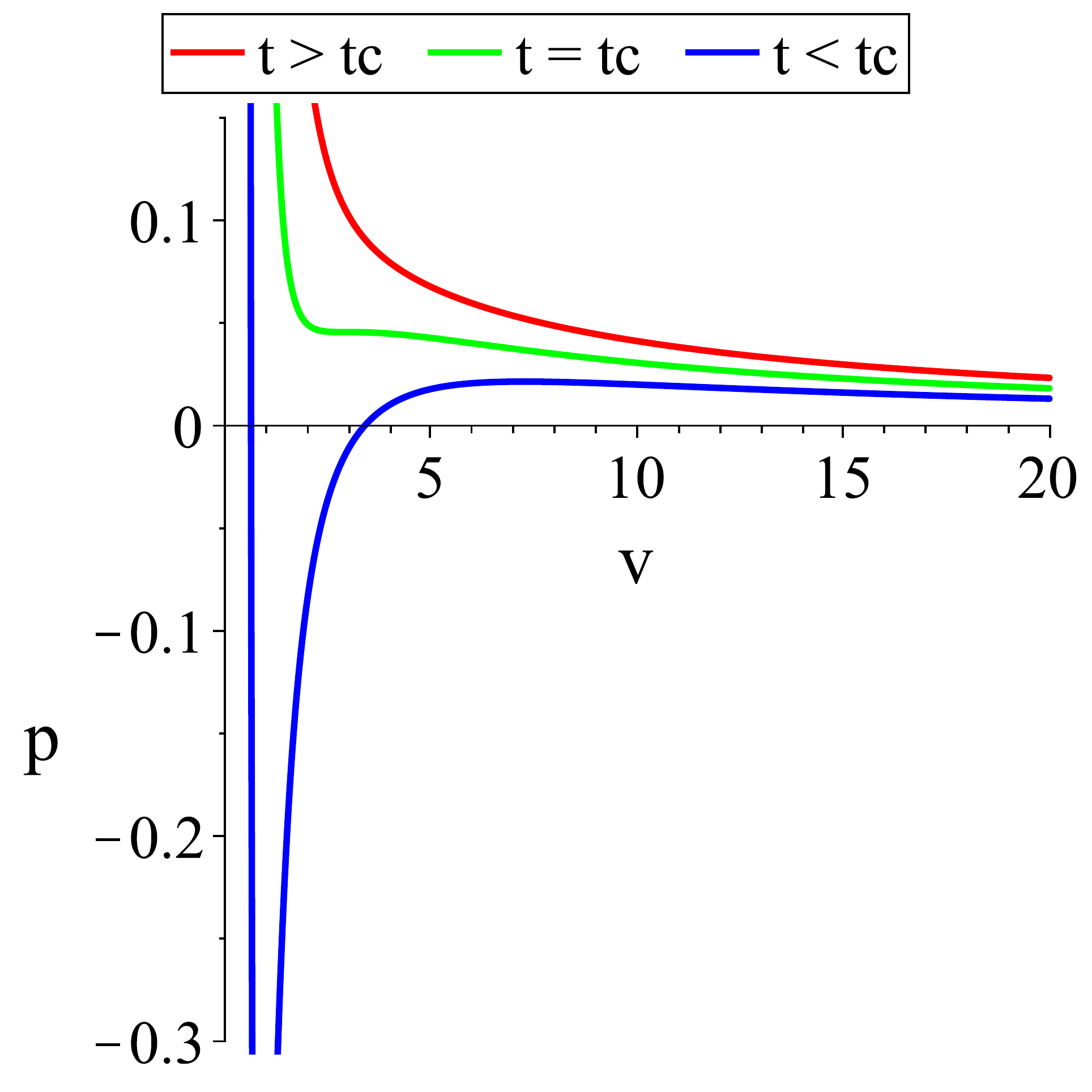}
    \includegraphics[width=0.325\textwidth]{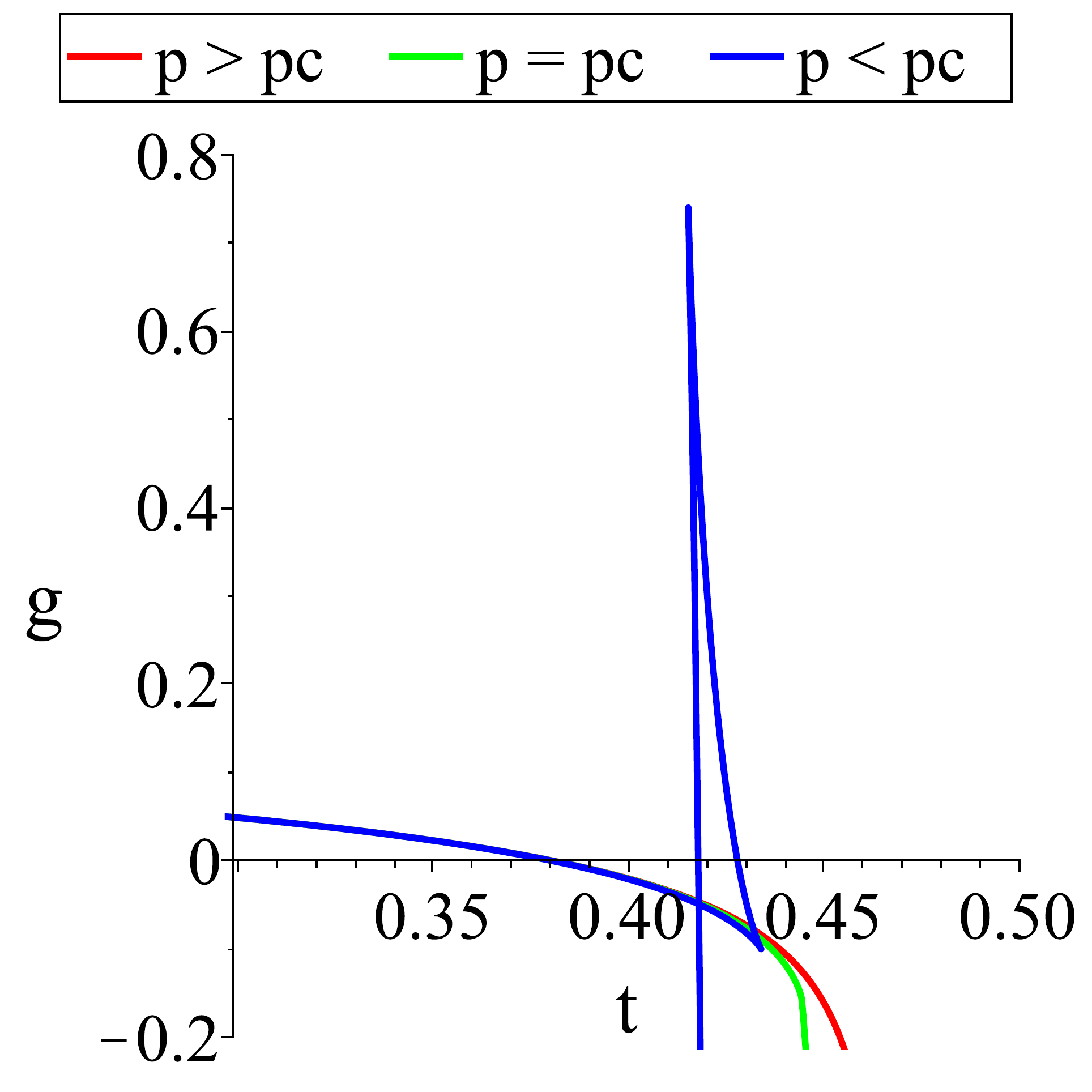}
    \includegraphics[width=0.325\textwidth]{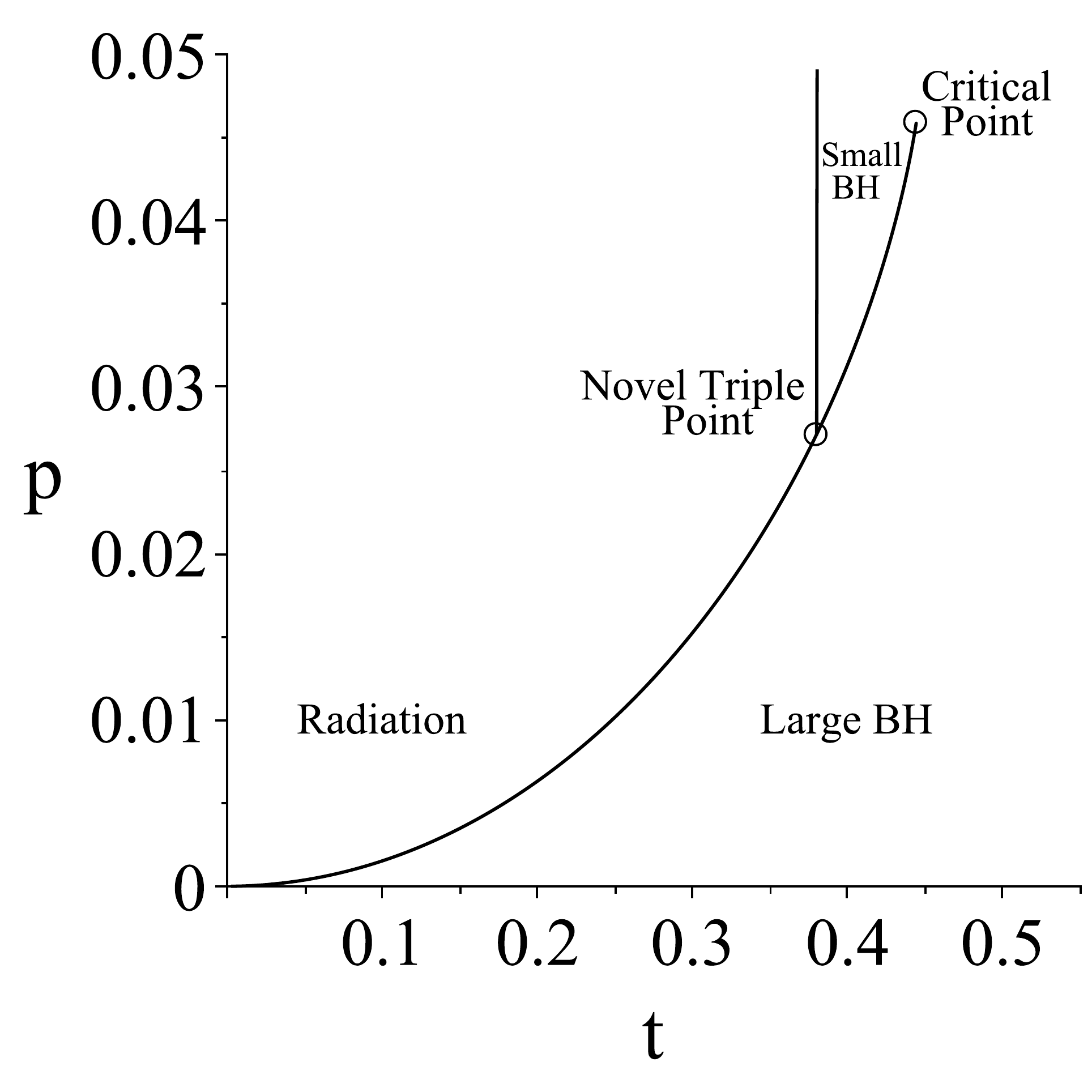}
    \caption{Seven-dimensional uncharged black holes with $b_{1}=b_{2}=b_{3}=1, \; \alpha=3$ and $q=0$. \textbf{Left:} $p\!-\!v$ isotherms with Van der Waals oscillations occurring for temperatures below $t_c$. \textbf{Middle:} $g\!-\!t$ isobars demonstrating the emergence of a swallowtail for pressures below $p_c$.  \textbf{Right:} Phase diagram, with coexistence lines (black) separating regions where the small, large, and radiation phases occur.}
    \label{pvgt7dalpha2b1}
\end{figure}

We begin with the uncharged solutions in seven dimensions. With $q=0$ the equation of state becomes
\begin{equation}
    p=\frac{t}{v}-\frac{5 b_{1}}{\pi v^{2}}+\frac{2 \alpha b_{1} t}{v^{3}}-\frac{5 \alpha }{2 \pi v^{4}}+\frac{3  t}{v^{5}}\ ,
\end{equation}
and critical points as determined by \eqref{criticalpoints} occur when
\begin{equation}
     t_{c}=\frac{10\left(b_{1} v_{c}^{2}+\alpha \right) v_{c}}{\pi\left(6 \alpha b_{1} v_{c}^{2}+v_{c}^{4}+15 \right)}\ ,\quad
 10 b_{1} v_{c}^{6}+60 \alpha^{2} b_{1} v_{c}^{2}-30 \alpha v_{c}^{4}-450 b_{1} v_{c}^{2}-150 \alpha=0\ .
\end{equation}
We first examine the case when $b_{1}=b_{2}=b_3=1$, where the above reduces to
\begin{equation}
    t_{c}=\frac{10\left(v_{c}^{2}+\alpha\right) v_{c}}{\pi\left(v_{c}^{4}+6 \alpha v_{c}^{2}+15\right)}\ , \quad 10 v_{c}^{6}-30 \alpha v_{c}^{4}-\left(450 - 60 \alpha^{2}\right) v_{c}^{2}-150 \alpha=0\ .
\end{equation}
The critical volume in seven dimensions is therefore given by the solution to a cubic function in $v_{c}^2$. Only one solution is real and positive for $\alpha \geq \sqrt{3}$.  Positivity of the entropy \eqref{entropy} will be automatically satisfied for black holes of any size since all coefficients are positive. In what follows, we fix the value of $\alpha$ to limit the size of the parameter space, noting where variations of $\alpha$ lead to important qualitative changes in the phase structure. Otherwise, changing $\alpha$ simply changes the numerical values of the various critical points and bounds we find herein. We set $\alpha=3$ so that we lie in the region of \figurename{ \ref{fig:alpharegion}} where all solutions for the metric function $f$ are real and positive, and a smooth Einstein limit exists. We are bounded only by a maximum pressure $p_+$, while the minimum pressure is null:
\begin{equation}
    p_+=\frac{5(4 \sqrt{6}-9)}{6 \pi} \approx 0.211\ , \quad p_-=0\ .
\end{equation}
In Figure~\ref{pvgt7dalpha2b1} we show the equation of state $p(v)$, the Gibbs free energy $g(t)$, and the phase diagram for these black holes. In  Figure~\ref{pvgt7dalpha2b1}a, the equation of state is shown for temperatures above, below, and equal to the critical temperature $t_c$. Below $t_c$, we observe Van der Waals oscillations characteristic of a small-large black hole phase transition. States in the negative pressure region are not only thermodynamically unstable (since $\partial p/\partial v>0$), but also correspond to asymptotically {\it de Sitter} configurations (since $\Lambda>0$). The physical state of the system instead follows the coexistence line as determined by Maxwell's equal area law, and is always positive in pressure due to the long tail of the $p(v)$ curve.
\\

In Figure~\ref{pvgt7dalpha2b1}b, we plot the Gibbs free energy as a function of equilibrium temperature at fixed pressure above, below, and at the critical pressure $p_c$, using $r_+$ as a parameter. The radiation phase (sometimes referred to as {\it thermal anti-de Sitter} or {\it empty anti-de sitter}) corresponds to the $g=0$ line. Along each curve, the black hole size increases as the curve is followed from left to right. As the temperature increases, the physical state of the system corresponds to whichever branch (including $g=0$) globally minimizes the free energy. Above the critical pressure (as in the red curve) we observe a Hawking-Page transition between the `empty' radiation phase and a large black hole where the red line crosses $g=0$. As the pressure decreases, a kink eventually forms at the critical pressure where a second-order phase transition between a small and large black hole occurs. Below $p_c$, a crossing forms where the small and large black hole branches are separated by an unstable intermediate branch, characterized by  a negative specific heat. As the temperature increases, the system follows the lower curve and a first-order small-large transition occurs since the horizon size is discontinuous at the crossing.
\\

\begin{figure}[H]
    \centering
    \includegraphics[width=0.3\textwidth]{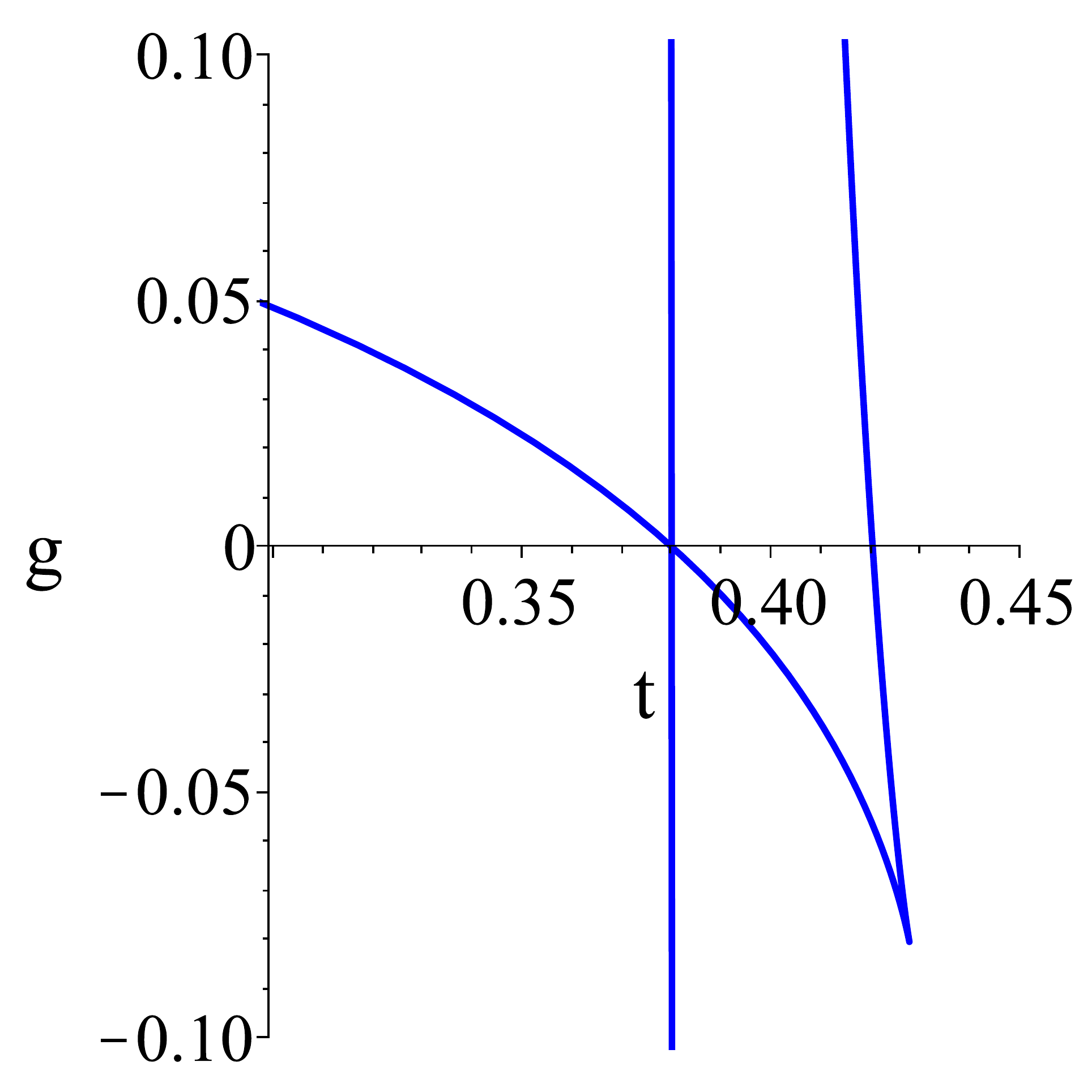}
    \caption{Free energy of the seven-dimensional uncharged black hole with $b_{1}=b_{2}=b_{3}=1, \; \alpha={3}$ and $q=0$. With $p\approx 0.027$, the small-large transition lies on the $g=0$ line, and a triple point occurs where the small and large black hole phases coexist with the radiation phase.}
    \label{7dgttripplot}
\end{figure}

A new feature appearing in our analysis is the presence of a {\it triple point} at sufficiently low pressure. As the pressure decreases towards zero the second-order phase transition occurring at the crossing within the swallowtail eventually intersects the radiation phase at $g=0$, as seen in Figure~\ref{7dgttripplot}. At this novel triple point, the large and small black hole phases coexist with the radiation phase, with all three globally minimizing $g$. This is in contrast to an ordinary triple point which consists of three phases that all correspond to black hole states. The vertical line in Figure~\ref{7dgttripplot} extends upwards and terminates at the maximum pressure $p_+$ (not shown). This novel triple point was also observed in Gauss-Bonnet black holes with non-constant curvature horizons \cite{hull2021}, and has not previously been noted in studies of the phase structure of anti-de Sitter black holes in third-order Lovelock gravity. Despite receiving significant attention in the hyperbolic ($\kappa=-1$) and flat ($\kappa=0$) cases, notably in \cite{farhangkhah2014}, the spherical $\kappa=1$ case has not been analyzed as thoroughly.

Finally, we observe that the phase structure is highly sensitive to variations of $\alpha$. As $\alpha$ increases, the critical pressure decreases, until $\alpha \approx 3.85$ where a turning point occurs and $p_c$ begins to increase monotonically with $\alpha$. When $\alpha \approx 4.56$ the critical pressure coincides with the maximum pressure $p_+$, while above this value, no critical points occur. Despite this, one can still examine phases with $\alpha\gtrsim4.56$, however in the uncharged case the phenomenology is limited to Hawking-Page phase transitions between radiation and a large black hole. When $\alpha \gtrsim 3.87$ the novel triple point no longer occurs at any allowed pressure. In this case the free energy at the small-large transition always lies above $g=0$, and only a Hawking-Page transition occurs (for any value of $p$). Finally as the value of $\alpha$ increases, the triple point moves to the right, eventually merging with the critical point. Above this value of $\alpha$ only a Hawking-Page transition occurs. Henceforth we fix $\alpha=3$, as variations in $\alpha$ simply correspond to a rescaling of the $b_n$ terms when $d\geq8$.

\subsection{Charged Solutions}
We now briefly examine charged solutions with positive-curvature transverse sections. When $q\neq0$ and $b_{1}=b_{2}=b_{3}=1$, the critical temperature and volume are given by
\begin{equation} \label{critVT}
  t_{c}={\frac {10{v_{{c}}}^{8}+10\alpha{v_{{c}}}^{6}-10\pi{q}^{2}}{\pi{v_{{c}}}^{5} \left( {v_{{c}}}^{4}+6\alpha{v_{{c}}}^{2}+15\right) }}\ ,
\end{equation}
\begin{equation}
 v_{c}^{12}-3\alpha v_{c}^{10}+ \left( 6{\alpha}^{2}-45
 \right) v_{c}^{8}-15\alpha{v_{{c}}}^{6}-(9\pi{v_{{c
}}}^{4}+42\pi\alpha{v_{{c}}}^{2}+75\pi){q}^{2}=0\ .
\end{equation}
The critical pressure is given by
\begin{equation}
   p_{c}={\frac {10{v_{{c}}}^{12}-5\alpha{v_{{c}}}^{10}+ \left( 10{
\alpha}^{2}-90 \right) {v_{{c}}}^{8}-15\alpha {v_{{c}}}^{6}-18\pi
{q}^{2}{v_{{c}}}^{4}-28\pi\alpha{q}^{2}{v_{{c}}}^{2}-30\pi
{q}^{2}}{2\pi{v_{{c}}}^{10} \left( {v_{{c}}}^{4}+6\alpha{v_{{c
}}}^{2}+15 \right) }}
\end{equation}
where $v_{c}$ will be the solution to the critical volume expression while the temperature has been replaced by the critical temperature. 
The critical volume equation can be re-expressed as the condition that $w_1=w_2$ where we define
\begin{equation}
w_{1}=\left(9 \pi v_{c}^{4}+42 \alpha \pi v_{c}^{2}+75 \pi\right) q^{2}, \quad w_{2}=v_{c}^{12}-3\alpha v_{c}^{10}+ \left( 6{\alpha}^{2}-45 \right) v_{c}^{8
}-15\alpha v_{c}^{6}
 .
\end{equation}
In this form the number of possible critical points can be easily determined through Descartes' rule of signs (graphically, we are concerned with the number of intersections between $w_1$ and $w_2$). $w_{1}$ is a quadratic in $v_{c}^2$ and no sign changes occur between terms, therefore no positive roots exist ($q$ acts simply as a rescaling factor). For $w_{2}$ there are three sign changes, so a maximum of three positive roots exist and a minimum of one. For the case considered here where $\alpha=3$ there is only one positive root, therefore for any value of $q$ there exists only one value of $v_{c}$ for which $w_{1}=w_{2}$, and only one critical point exists. In what follows the charge is fixed to $q=1$ unless charge variations are being considered.
\\

\begin{figure}[H]
    \includegraphics[width=0.325\textwidth]{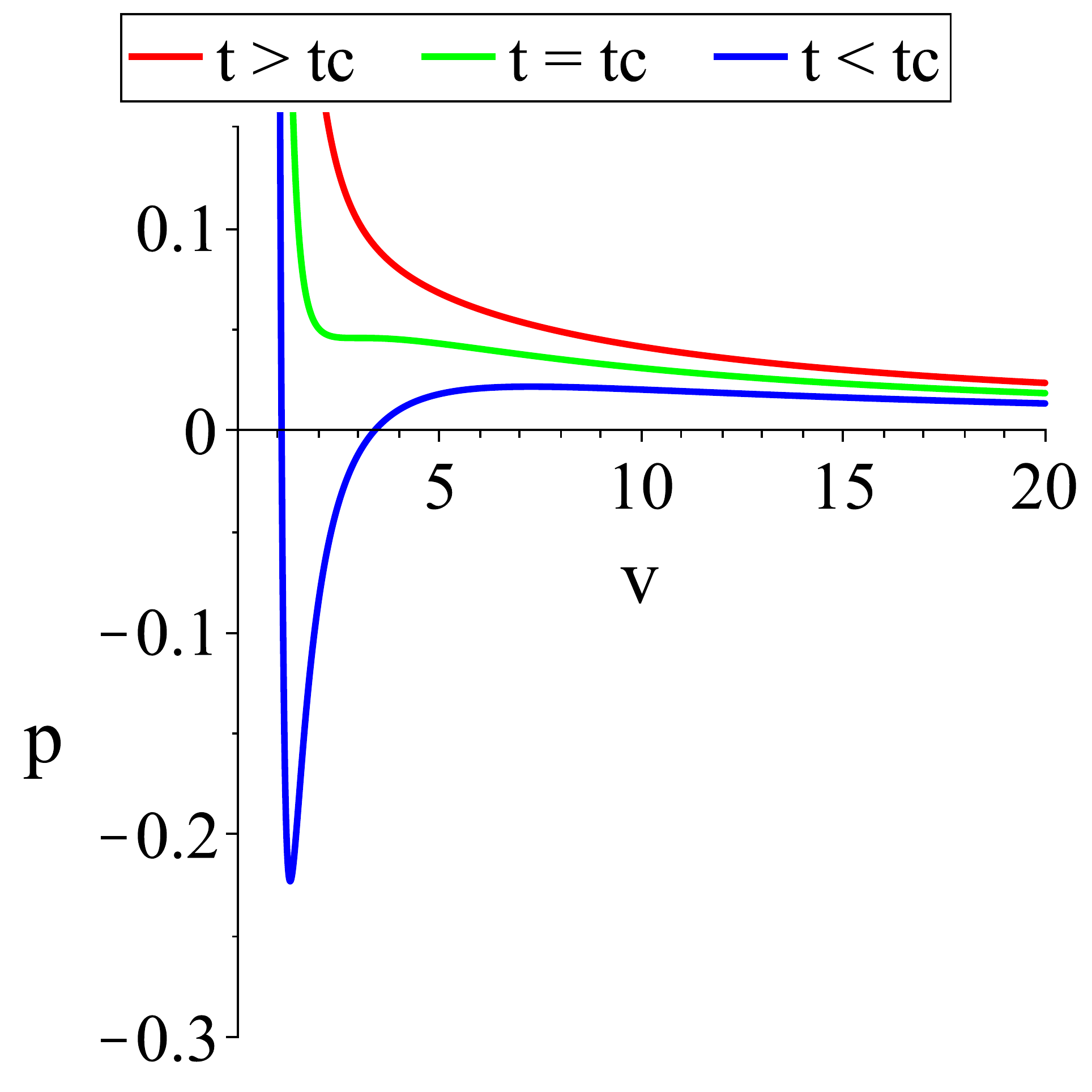}
    \includegraphics[width=0.325\textwidth]{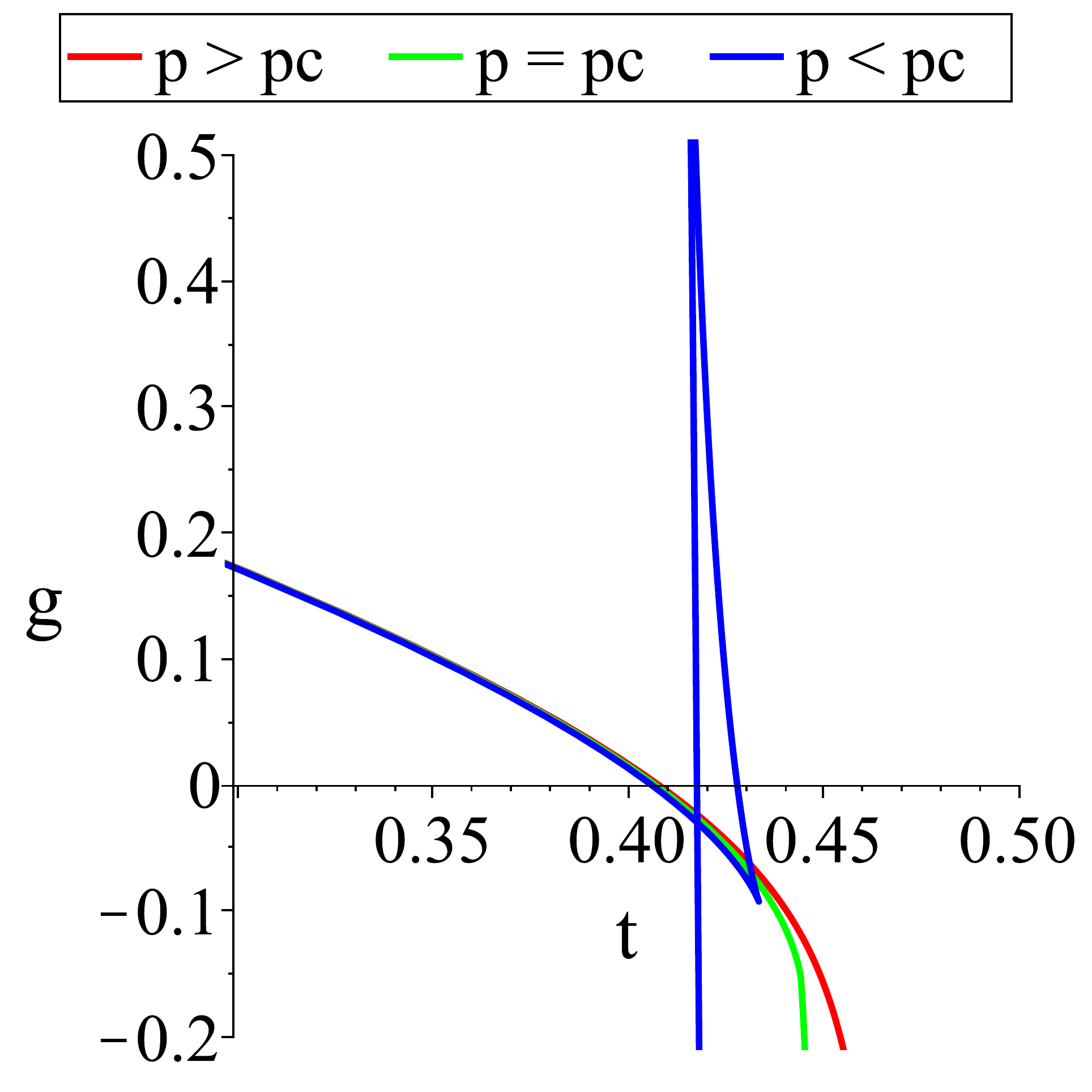}
    \includegraphics[width=0.325\textwidth]{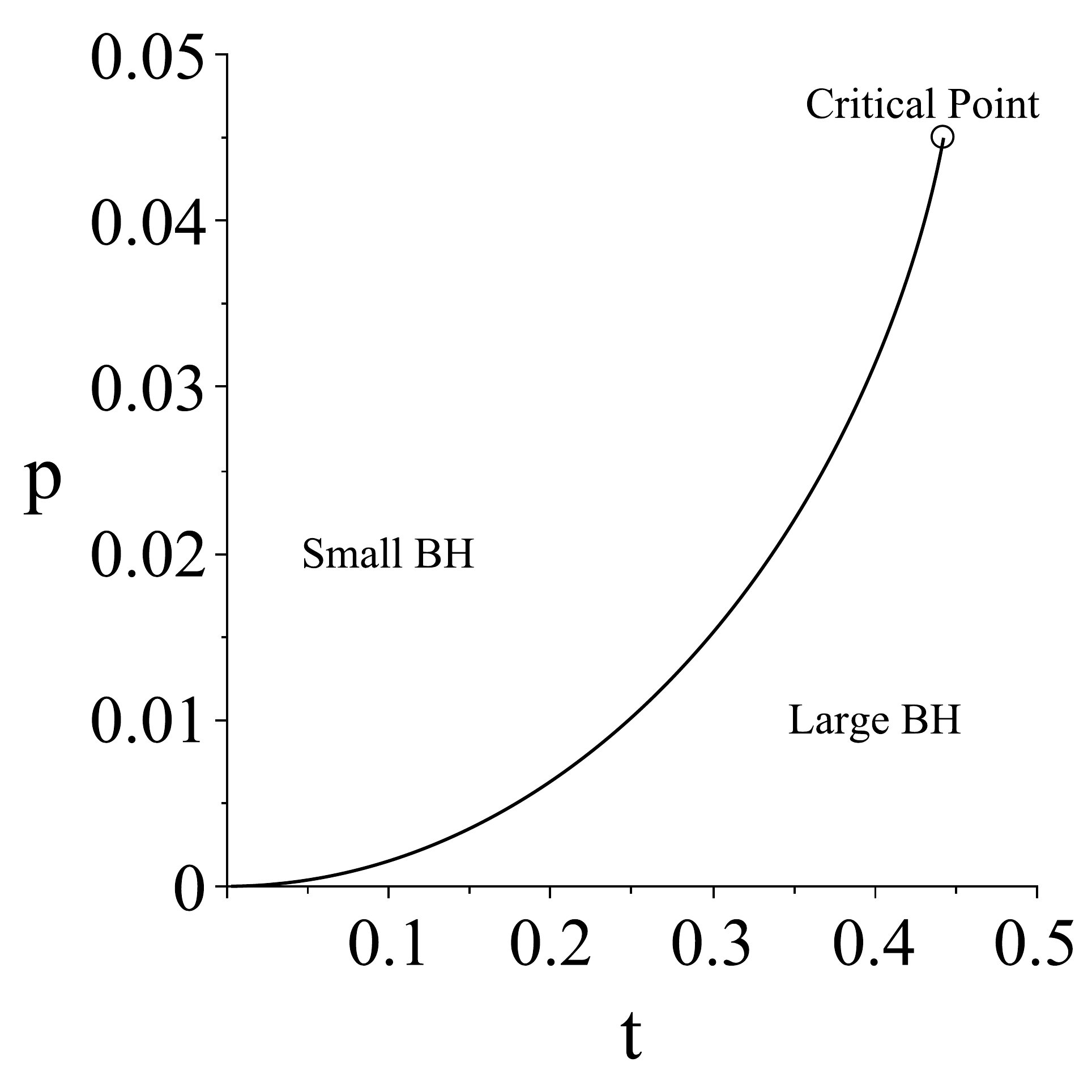}
    \caption{Seven-dimensional charged black holes with $b_{1}=b_{2}=b_{3}=1$, $ \alpha=3$, and $q=1$.\hspace{5pt} \textbf{Left:} $p\!-\!v$ isotherms with Van der Waals oscillations occurring for temperatures below $t_c$. \textbf{Middle:} $g\!-\!t$ isobars demonstrating the emergence of a swallowtail for pressures below $p_c$.  \textbf{Right:} Phase diagram, with coexistence lines (black) separating regions where the small and large black hole phases occur.}
    \label{7dchargedb11}
\end{figure}

As with the uncharged solutions, we examine the thermodynamics behaviour through the equation of state and free energy, as shown in Figure~\ref{7dchargedb11}. On the right, the same oscillatory behaviour seen in the uncharged case is present, though the transition to the oscillatory region is now sensitive to the charge. As the value of charge is increased, the critical temperature decreases. In Figure~\ref{7dchargedb11}b, the behaviour of the Gibbs free energy is qualitatively similar to the uncharged case, but with a different interpretation. In the canonical ensemble the physical charge is fixed, which places a lower bound on the allowed black hole size. In the case of $d=4$ Einstein gravity for example, Reissner-N\"{o}rdstrom black holes require that $m^2>q^2$ otherwise a naked singularity forms. This constraint generalizes to higher-dimensional Lovelock black holes, placing a lower bound on the black hole size which depends on $\{\alpha,q,\Lambda\}$. Thus, Hawking-Page transitions do not occur since the radiation phase corresponds to an $m=0$ black hole which cannot exist when $q=0$.  Instead, we observe a first-order phase transition between a small and large black hole when $p<p_c$, a second-order phase transition when $p=p_c$, and no transitions when $p>p_c$.

Before moving on to eight-dimensions, we make a brief statement about the hyperbolic case. When $b_{1}=-1,b_{2}=1,b_{3}=-1$, there are no  critical points for $\alpha=3$. Critical points are possible for values of $\alpha<3$ as seen in \cite{frassino2014}, however these values of $\alpha$ appear to correspond to spacetimes where there is no smooth limit between the three branches of solutions to \eqref{fpoly}, hence their properties and structures are not entirely understood. Furthermore,  hyperbolic geometries contain discontinuities in both the Gibbs free energy and temperature making the thermodynamic analysis more difficult. We therefore limit our analysis to $b_1>0$ where such issues are not present.

\section{Thermodynamics of $d=8$ Exotic Black Holes} \label{8d}

In eight dimensions, we no longer require that $b_{1}=\kappa$, $b_{2}=\kappa^{2}=1$ and $b_{3}=\kappa^{3}$. However some restrictions can be placed on the $b_n$ by considering the character of the resulting metric functions. In the exotic black hole solutions given by \eqref{fpoly}, $b_1$ can in principle be chosen arbitrarily. However when $b_1<0$, any would-be Hawking-Page transitions involve a change in topology in the asymptotic region, since the spatial sections are hyperbolic in the black hole spacetime but flat in the `empty' AdS spacetime. This represents a topological transition in addition to the thermal Hawking-Page transition. Though in a quantum setting such transitions may be possible, we will avoid these cases so as to isolate the purely thermal features of the phase structure. Moreover, having $0\!<b_1\!<1$ or $b_1\!>\!1$ is also possible, but such a choice introduces a conical deficit (or excess) in the spacetime. The metric is no longer Ricci flat, instead resembling metrics that arise in the presence of a global monopole charge \cite{vilenkin1994}. In the simplest case, such metrics are sourced by $\mathcal{O}(3)$ scalar fields with Higgs potentials whose vacuum expectation value does not vanish. Being interested in vacuum solutions with no external fields, we wish to avoid these cases as well. As a result, we will restrict our attention to $b_1=1$ throughout. $b_2$ and $b_3$ are not constrained in this way, as they do not determine the asymptotic structure of the spacetime.
\\

\begin{figure}[h]
\includegraphics[width=0.5\textwidth]{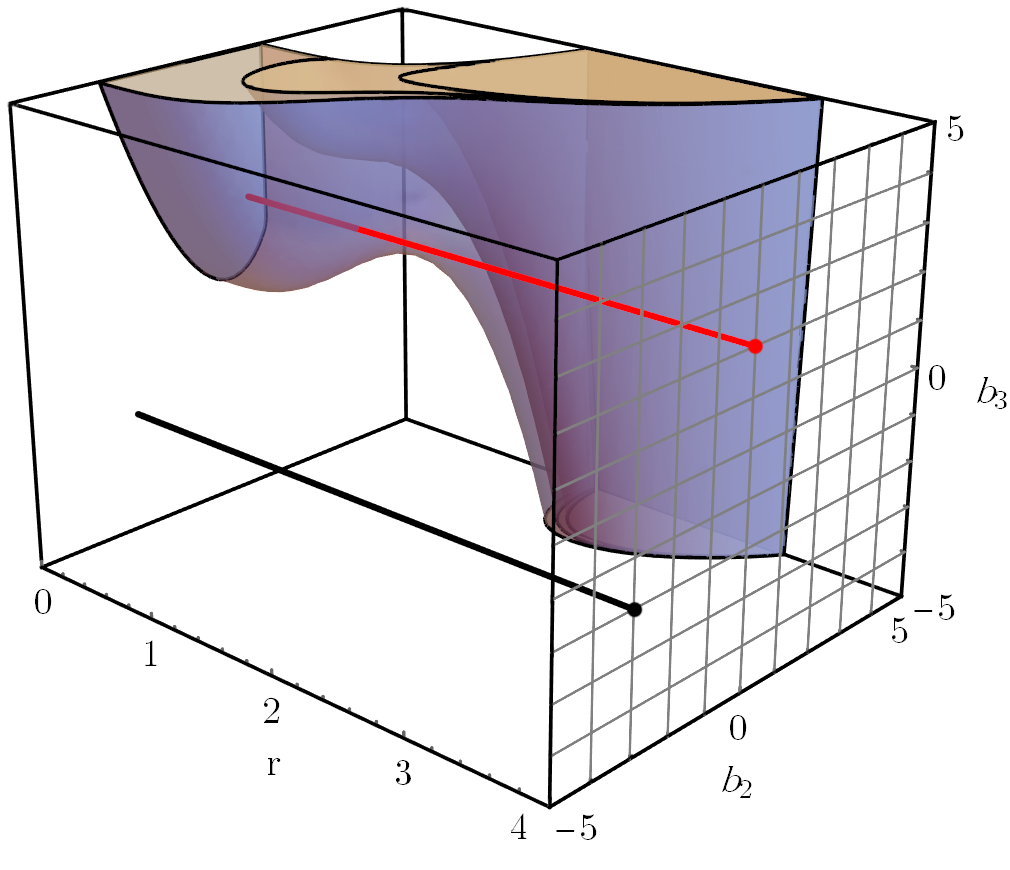}\quad \includegraphics[width=0.4\textwidth]{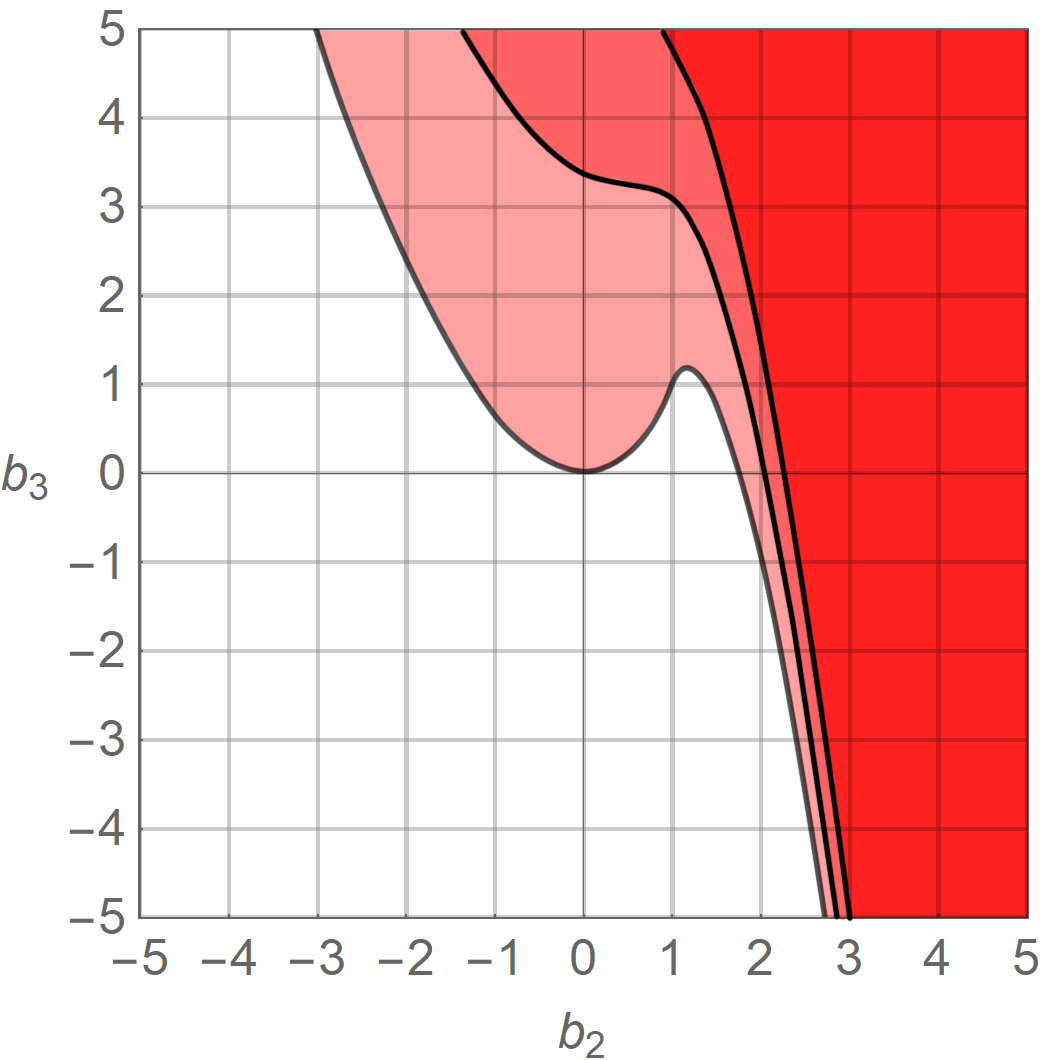}
\caption{Inadmissible choices of the parameters $\{b_n\}$ for the 8-dimensional AdS geometry, with $m\in\{0,50,100\}$, $b_1=1$, $\alpha_0=0.01$ and $\alpha=3$. {\bf Left:} Within the shaded region, $f(r)$ is not defined, while on the boundary of the shaded region $\mathcal{K}$ diverges.  {\bf Right:} Projection onto the $b_2$--$b_3$ plane, with the exclusion regions shaded in red.}
\label{fig:kret}
\end{figure}

Further restrictions can be placed on $b_2$ and $b_3$ however by considering the behaviour of the Kretschmann invariant $\mathcal{K}=R_{abcd}R^{abcd}$. In $d$-dimensions, its functional form is given by
\begin{equation}\label{kret1}
\mathcal{K}=\left(\frac{d^{2} f(r)}{d r^2}\right)^{2} +  \frac{2(d-2)}{r^2}\left( \frac{d f(r)}{dr} \right) ^2 + \frac{2(d-2)(d-3) f(r)^2}{r^4}  - \frac{4 R[\gamma] f(r)}{r^4} +\frac{\mathcal{K}[\gamma]}{r^4}\ ,
\end{equation}
whereas before $f(r)$ corresponds to the chosen branch of \eqref{fpoly}, $R[\gamma]$ is the scalar curvature of the base manifold, and $\mathcal{K}[\gamma]$ is the Kretschmann scalar of the base manifold. $\mathcal{K}$ will diverge not only when $f(r)$ or its derivatives diverge, but also when the following quantity vanishes
\begin{align}\label{x1}
X(r)\equiv&\  81\, b_2-27\, b_3-54+ \left(9\alpha -2 \alpha^{3} -27 \alpha_0\right) r^{6}\nonumber\\
&+\sqrt{4\big(9\, b_2-9-\left(\alpha^{2}-3\right) r^{4}\big)^{3}-\big(81\, b_2-27\, b_3-54+\left(9 \alpha-2 \alpha^{3}-27 \alpha_0\right) r^{6}\big)^{2}}\ ,
\end{align}
which can be seen by direct substitution of \eqref{ebranch} into \eqref{kret1}. For all reasonable\footnote{Reasonable here means real-valued and at most $\mathcal{O}(1)$.} choices of couplings and $b_n$ the metric function $f(r)$ is $\mathcal{C}^{\infty}$, so one need only avoid choices of parameters for which $X(r)$ vanishes for some $r$. \figurename{ \ref{fig:kret}} shows regions in the $\{b_n\}$ parameter space where this occurs for $\alpha_0=0.01$ and $\alpha=3$. 

On the left of \figurename{ \ref{fig:kret}}, we show inadmissible regions of the parameter space. The metric function is not defined within the shaded region, while on the boundary surface the Kretschmann invariant diverges\footnote{We have excluded the divergence of $K$ at $r=0$ which occurs generically for any choice of base manifold, couplings, and $b_n$'s.}. We show two choices of $b_2$ and $b_3$ (in black and red) demonstrating a choice where the curvature remains finite for all $r$ (black) and one where the curvature diverges at finite $r$ (red). In the general case the singularities represented by the boundary surface are not hidden behind an accompanying event horizon. On the right of \figurename{ \ref{fig:kret}}, we show permissible choices of $b_n$ by projecting out the radial dimension from the left figure. The shaded region then indicates that $\mathcal{K}$ diverges at {\it some} finite $r$ for the given choice of $b_2$ and $b_3$. It should be noted that for the $m=0$ case, the black boundary asymptotically approaches the $b_3=0$ line, but does not intersect it. Therefore, the case where $b_2=b_3=0$ is not pathological. It is also important to note the mass-dependence of the inadmissible region. One may appear to have, for example, a Hawking-Page transition from a black hole with $b_2=0$, $b_3=1$ and $m=50$ to an empty spacetime, but such a transition is not allowed since when $m=0$ (as in the radiation phase) one cannot actually have $b_2=0$ and $b_3=1$.

In what follows, we restrict our analysis to choices that lie outside of this region, thereby avoiding parameters for which naked singularities may form.

\subsection{Uncharged Solutions}
In $d=8$ with $q=0$ and $\alpha=3$, the maximum and minimum pressures are
\begin{equation}
    p_{+}=\frac{7(-27+12 \sqrt{6})}{18 \pi} \approx 0.296\ ,\quad p_-=0\ ,
\end{equation}
and the equation of state becomes
\begin{equation}
    p=\frac{t}{v}-\frac{15}{2 \pi v^{2}}+\frac{6 t}{v^{3}}-\frac{27 b_{2}}{2 \pi v^{4}}+\frac{3 b_{2} t}{v^{5}}-\frac{3 b_{3}}{2 \pi v^{6}}\ .
\end{equation}
The resulting critical volume and temperature are then given by
\begin{gather}
    t_{c}=\frac{3\left(5 v_{c}^{4}+18 b_{2} v_{c}^{2}+3 b_{3}\right)}{\pi v_{c}\left(v_{c}^{4}+18 v_{c}^{2}+15 b_{2}\right)}\ , \\
    15 v_{c}^{8}+\left(162 b_{2}-270\right) v_{c}^{6}+\left(297 b_{2}+45 b_{3}\right) v_{c}^{4}-\left(810 b_{2}^{2}-486 b_{3}\right) v_{c}^{2}+135 b_{2} b_{3}=0\ . \label{vc8d}
\end{gather}
While the topological parameters are independent of each other, thermodynamic considerations lead to further constraints on their values. Re-arranging the equation of state \eqref{eos1} as
\begin{equation}
    t\rightarrow t(p,v,b_{2},b_{3})=\frac{2 \pi p v^{6}+15 v^{4}+27 v^{2} b_{2}+3 b_{3}}{2 \pi v\left(v^{4}+6 v^{2}+3 b_{2}\right)}\ 
\end{equation}
reveals that negative values of $b_{2}$ lead to a divergent temperature due to the $(v^{4}+6 v^{2}+3 b_{2})$ term in the denominator. We therefore restrict our analysis to the case where $b_{2} \geq 0$. This $(v^{4}+6 v^{2}+3 b_{2})$ term is also present in the free energy $g$, which then diverges along with the temperature. From equation \eqref{vc8d} we can determine the possible number of positive real $v_{c}$ solutions through Descartes' rule of signs, which are summarized in Table~\ref{tab} below.
\begin{table}[H]
\centering
\begin{tabular}{|c| c |c |} 
 \hline
  & $b_{2}=0$ & $b_{2}>0$ \\
 \hline
 $b_{3}=0$ & 1 & 3,1  \\ 
 \hline
 $b_{3}<0$ & 1 & 3,1  \\
 \hline
 $b_{3}>0$ & 2,0 & 4,2,0  \\
 \hline
\end{tabular}
\caption{Possible number of positive real roots to \ref{vc8d} from Descartes' rule of signs applied to the topological parameters $b_{2}$ and $b_{3}$.}
\label{tab}
\end{table}

\begin{figure}[H]
    \includegraphics[width=0.325\textwidth]{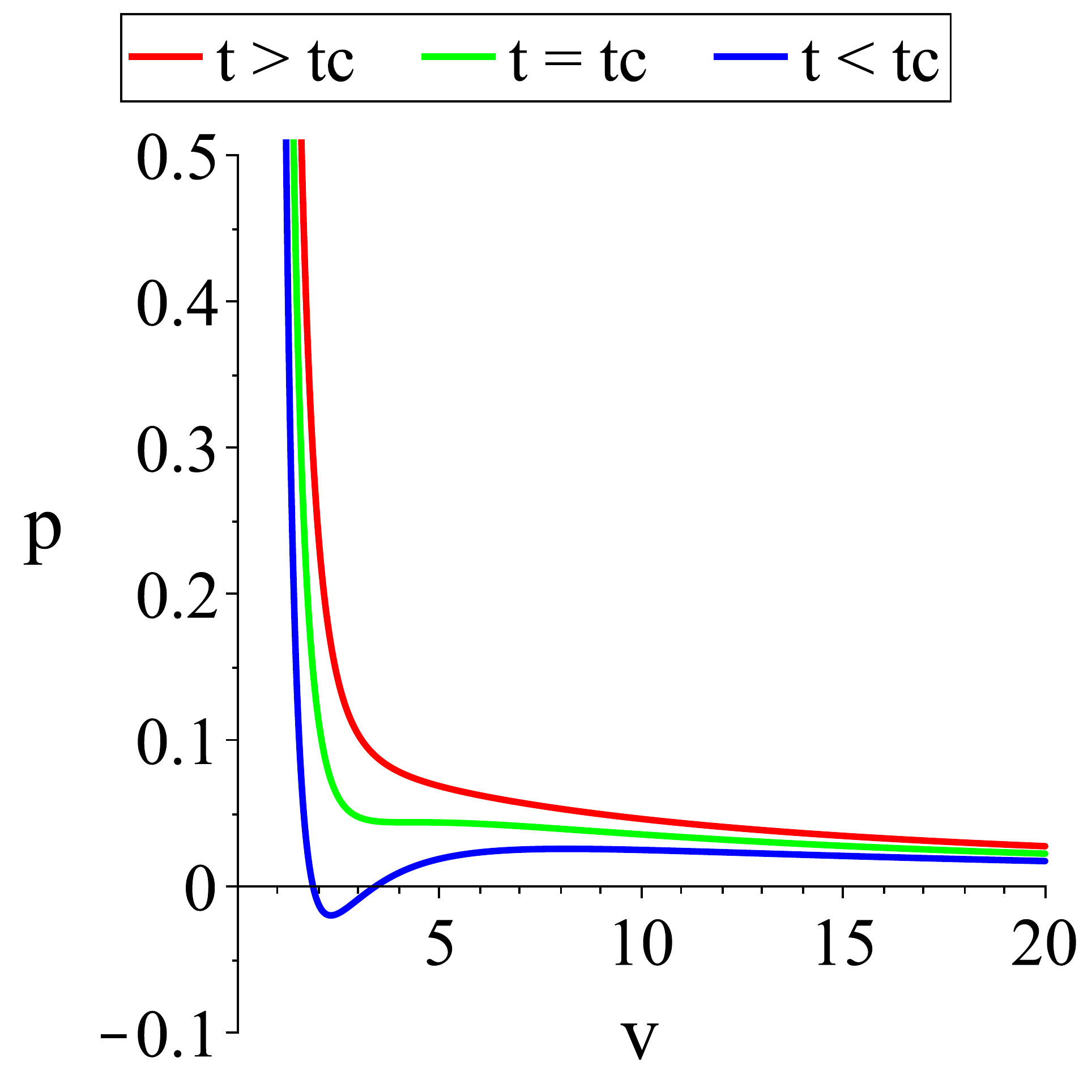}
    \includegraphics[width=0.325\textwidth]{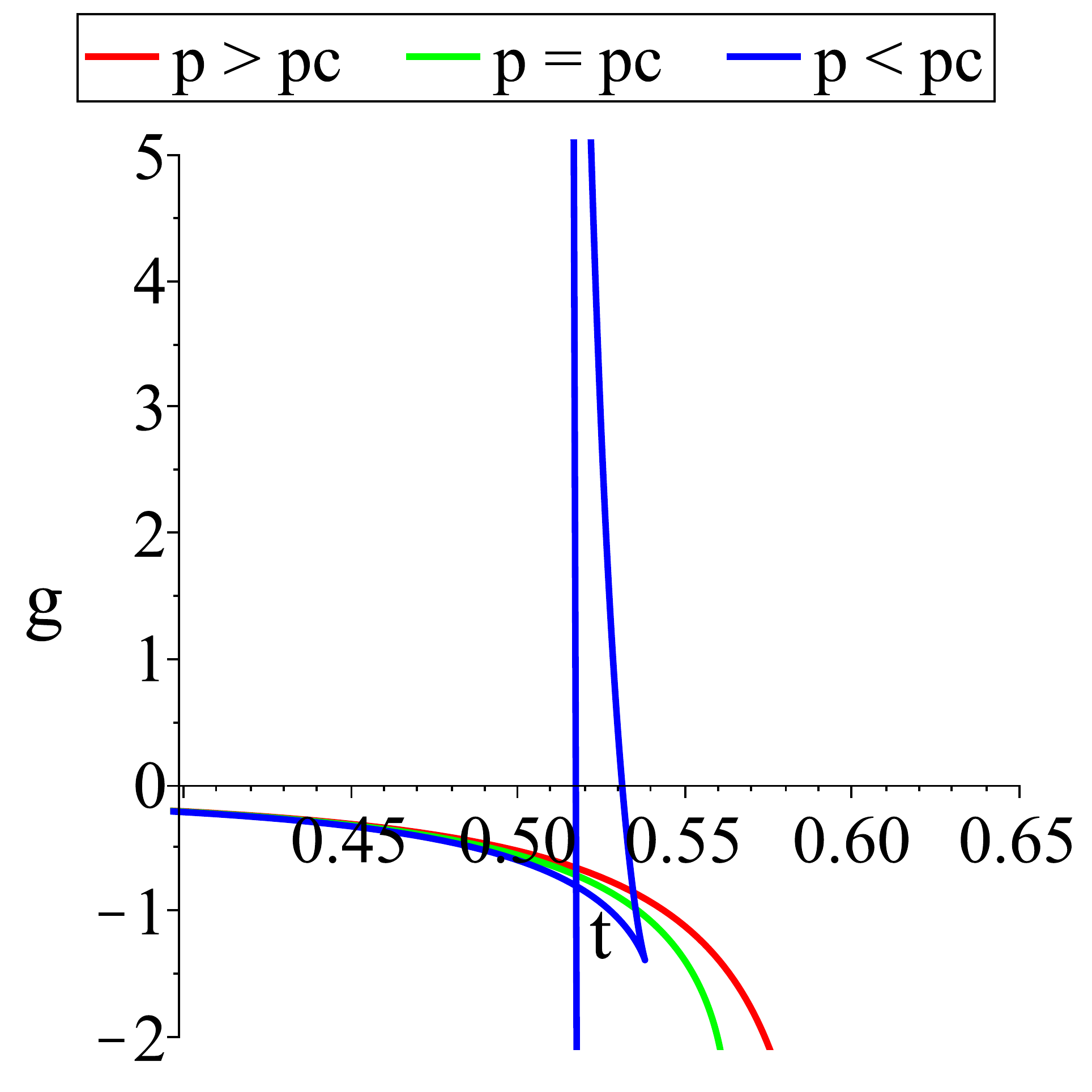}
    \includegraphics[width=0.325\textwidth]{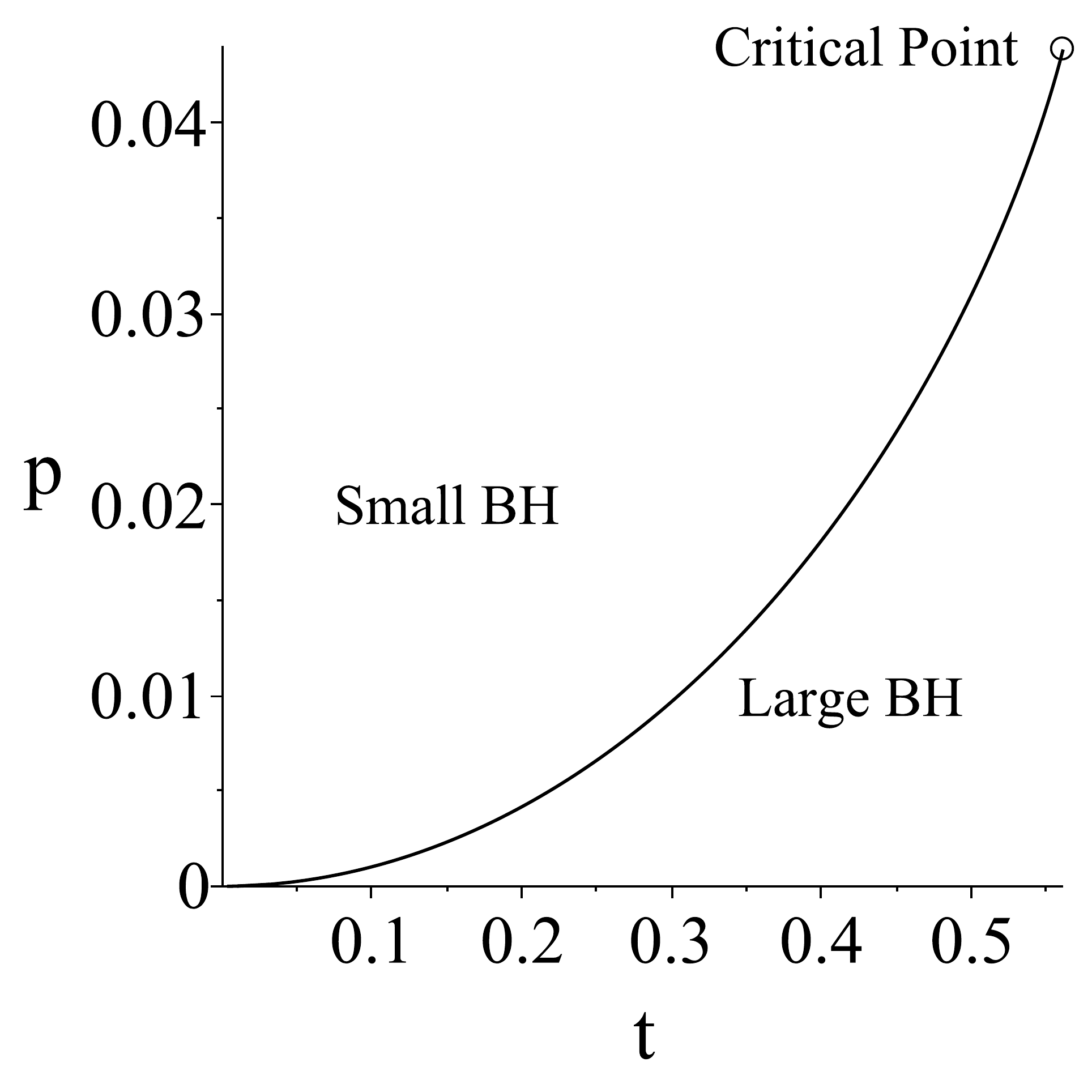}
    \caption{Eight-dimensional uncharged black holes with $b_{1}=1$, $b_{2}=0$, $b_{3}=-1$, and $\alpha=3$. \textbf{Left:} $p\!-\!v$ isotherms with Van der Waals oscillations occurring for temperatures below $t_c$. \textbf{Center:} $g\!-\!t$ isobars demonstrating the emergence of a swallowtail for pressures below $p_c$.  \textbf{Right:} Phase diagram, with coexistence lines (black) separating regions where the small and large black hole phases occur.}
    \label{b20b3-1}
\end{figure}

We begin with the case where $b_{2}=0$ and $b_{3}=-1$. This choice results in thermodynamic behaviour similar to the seven-dimensional case, with small-large transitions appearing generically below the critical temperature and pressure, as shown in Figure~\ref{b20b3-1}. Note that there is no Hawking-Page transition at low temperatures, and that $G\rightarrow 0$ as $T\rightarrow 0$.
\\

Criticality is also present when $b_{2}=0$ and $b_{3}=0$. The free energy in this case is qualitatively similar to that of Figure~\ref{b20b3-1}b, and the $p-v$ curves likewise follow Figure~\ref{b20b3-1}a. The critical pressure and temperature both increase as $b_3$ increases. 

When $b_{2}=0$ and $b_{3}>0$, two real and positive solutions for $v_{c}$ will exist for $b_{3}<\frac{324}{25}$. When $b_{3}\geq 1$, the novel triple point emerges (seen also in the uncharged seven-dimensional case above) with notable qualitative differences. Increasing $b_{3}$ likewise increases the temperature and pressure at which this triple point occurs.  Eventually, when $b_{3} \approx 6$, the triple point is no longer present as the novel triple point merges with the smaller of the two critical points and only the Hawking-Page transition is seen.

Next we consider the case where $b_{2}\neq 0$.  When $b_{2}=1$ and $b_{3}=0$, Van der Waals behaviour is observed corresponding to a small-large black hole transition, though no triple point occurs at any temperature/pressure. When $b_{3}>0$ instead, we observe the same qualitative behaviour noted above for $b_{2}=0$ and $b_3>0$: the existence of a triple point below a certain value of the parameter (in this case $0<b_{3} < 0.8$) , and Hawking-Page transitions when $b_3>0.8$.

\subsection{Charged Solutions}

When $d=8$ and $q\neq0$ the equation of state \eqref{eos1} becomes
\begin{equation}
    p=\frac{t}{v}-\frac{15}{2 \pi v^{2}}+\frac{6 t}{v^{3}}-\frac{27 b_{2}}{2 \pi v^{4}}+\frac{3 b_{2} t}{v^{5}}-\frac{3 b_{3}}{2 \pi v^{6}}+\frac{q^{2}}{v^{12}}\ ,
\end{equation}
with the critical temperature and volume given by
\begin{equation}
        t_{c}=-\frac{3\left(-5 v_{c}^{10}-18 b_{2} v_{c}^{8}-3 b_{3} v_{c}^{6}+4 \pi q^{2}\right)}{\pi v_{c}^{7}\left(v_{c}^{4}+18 v_{c}^{2}+15 b_{2}\right)}\ ,
\end{equation}
\begin{align}\label{critvol}
   15 v_{c}^{14}+\left(162 b_{2}-270\right) v_{c}^{12}+\left(297 b_{2}+45 b_{3}\right) v_{c}^{10}-\left(810 b_{2}^{2}-486 b_{3}\right) v_{c}^{8}+135 b_{2} b_{3} v_{c}^{6}\nonumber\\
   -\left(132 \pi v_{c}^{4}+1944 \pi v_{c}^{2}+1260 \pi b_{2}\right) q^{2}=0
\end{align}

In principle, a systematic determination of the root structure of the polynomial above can be done using Descartes' rule of signs; however, the presence of $b_3$ and $q$-dependent terms means such an analysis quickly becomes cumbersome. Instead we proceed on a case-by-case basis, as only a few examples are required to demonstrate the salient features of charged $d=8$ black holes. 

\begin{figure}[H]
    \centering
    \includegraphics[width=0.32\textwidth]{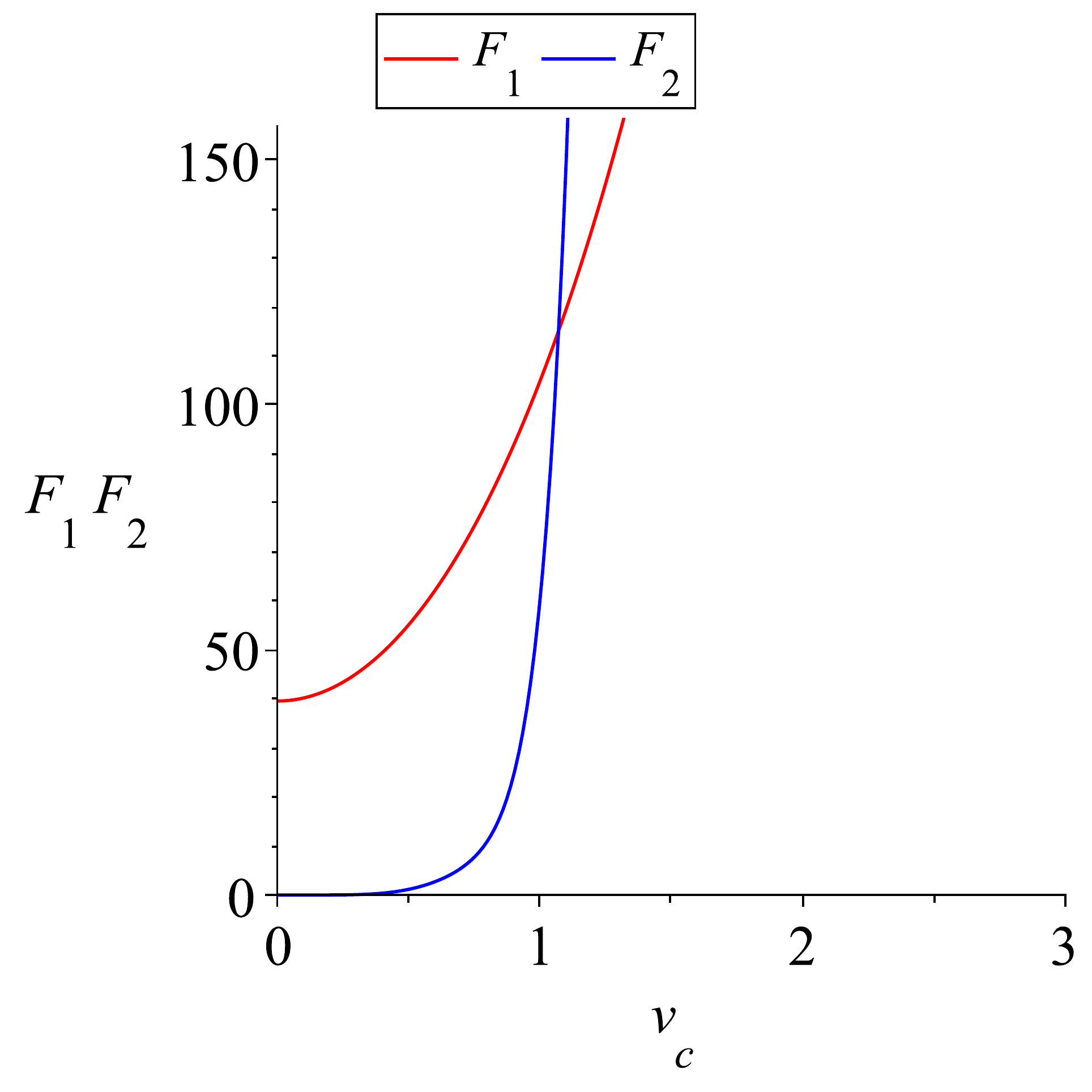}
    \includegraphics[width=0.32\textwidth]{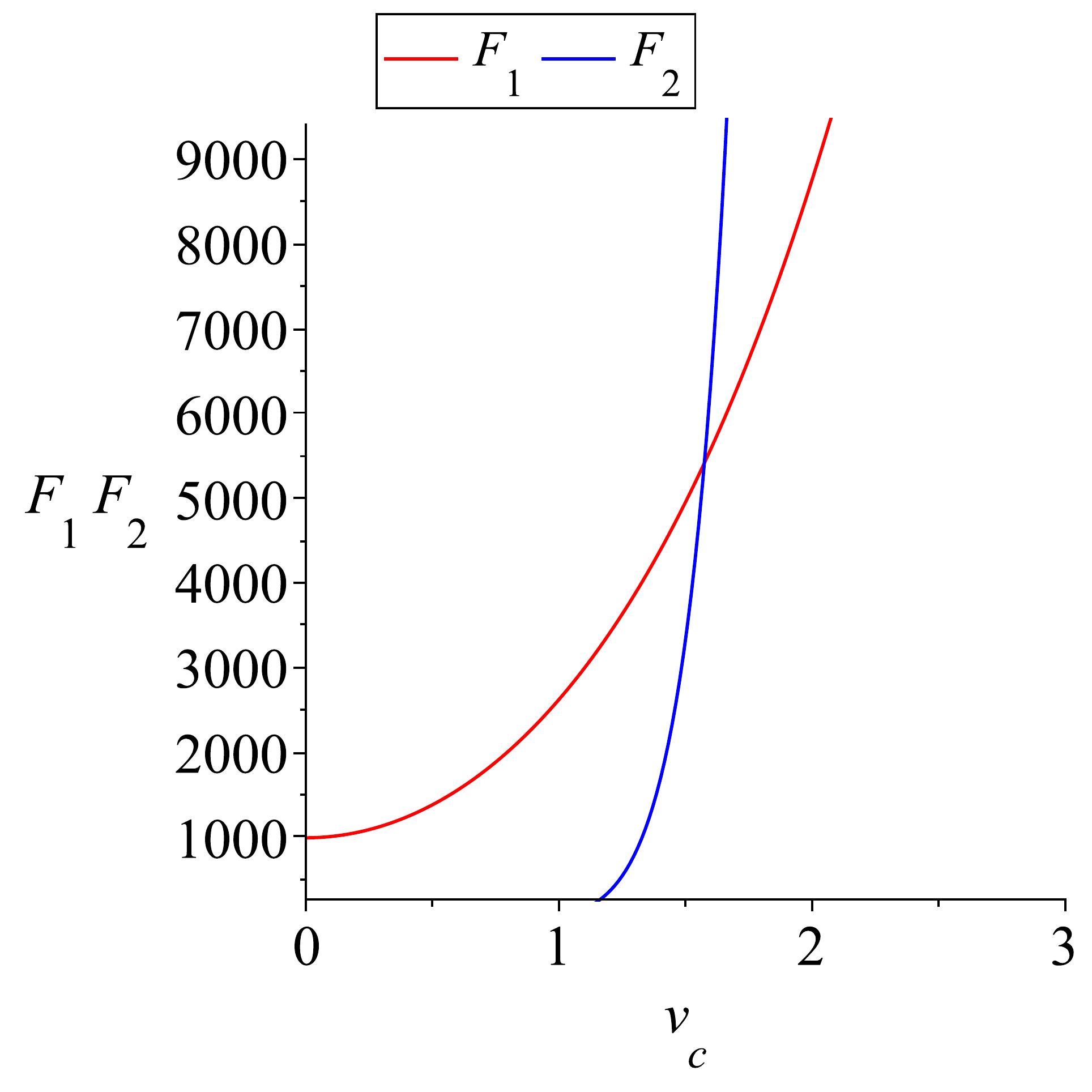}
    \includegraphics[width=0.32\textwidth]{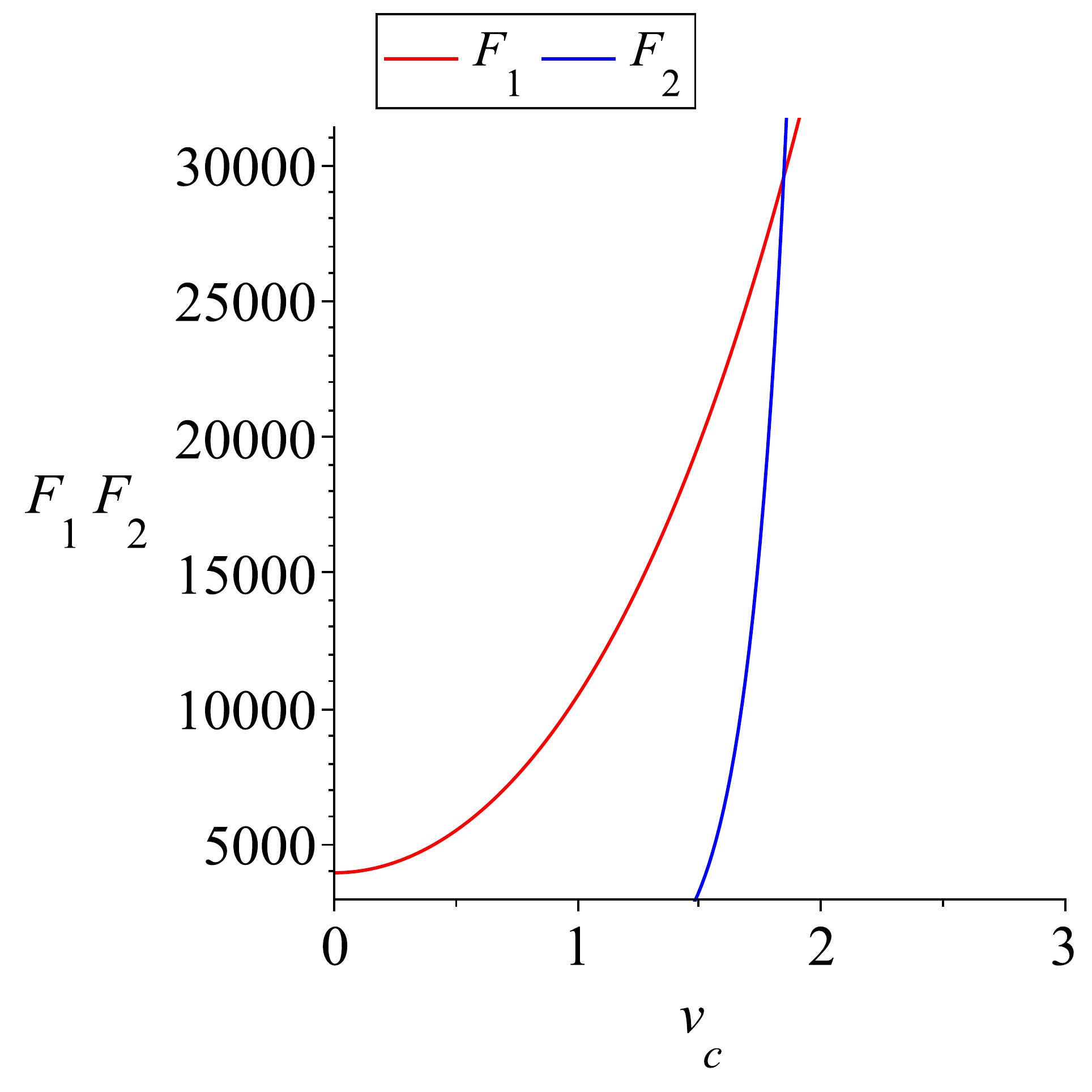}
    \caption{Plots for $F_{1}$ and $F_{2}$ as a function of $v_{c}$ for varying values of $q$. \textbf{Left:} $q=0.1$, \textbf{Center:} $q=0.5$, \textbf{Right:} $q=1$.}
    \label{Intersect}
\end{figure}

As with the uncharged case, we begin with the \textit{spherical} base manifold for which $b_{n}=1$. In this case, we observe only a small-large transition terminating at a critical point $\{t_c,p_c\}$ for all values of the charge $q$. This behaviour is identical to what is shown in Figure~\ref{7dchargedb11} for the $d=7$ case. As $q$ increases, $t_c$ and $p_c$ decrease, until they eventually achieve the $q=0$ values shown in Figure~\ref{b20b3-1}. That there can {\it only} be a small-large transition for 8-dimensional charged black holes is easily seen by separating the charge-dependent and charge-independent portions of \eqref{critvol}, and studying the number of intersections of the resulting functions. Let
\begin{equation}\label{f1f2}
    F_1=F_2
\end{equation}
with
\begin{equation}
    \begin{aligned}
    F_{1}&\equiv\left(132 \pi v_{c}^{4}+1944 \pi v_{c}^{2}+1260 \pi\right) q^{2}\\
    F_{2}&\equiv15 v_{c}^{14}-108 v_{c}^{12}+342 v_{c}^{10}-324 v_{c}^{8}+135 v_{c}^{6}\ .
    \end{aligned}
\end{equation}
Plotting $F_1$ and $F_2$ for varying values of $q$ reveals that for any finite value of the charge there is a unique value of $v_c$ which satisfies \eqref{f1f2}, as displayed in Figure~\ref{Intersect}.

As in the uncharged case, we plot the equation of state $p(v)$ and free energy $g(t)$, searching for Van der Waals oscillations in the $p-v$ plane and the formation of a swallowtail in the $g-t$ plane. Note that even though critical points may occur above $g=0$, the radiation phase is inaccessible to the system since the charge is fixed in the canonical ensemble, so no Hawking-Page transitions will occur in any of the cases we examine below.

\begin{figure}[H]
    \includegraphics[width=0.325\textwidth]{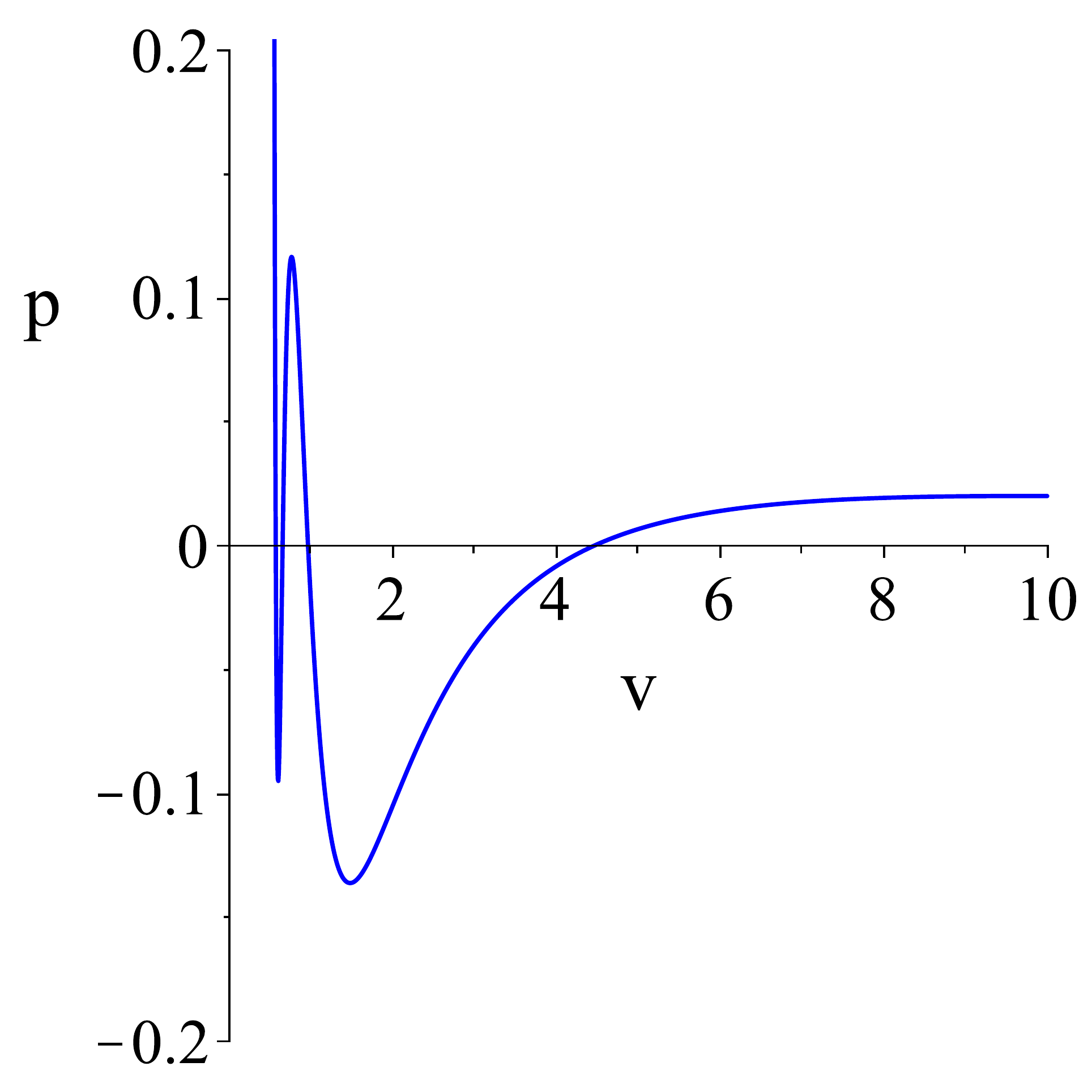}
    \includegraphics[width=0.325\textwidth]{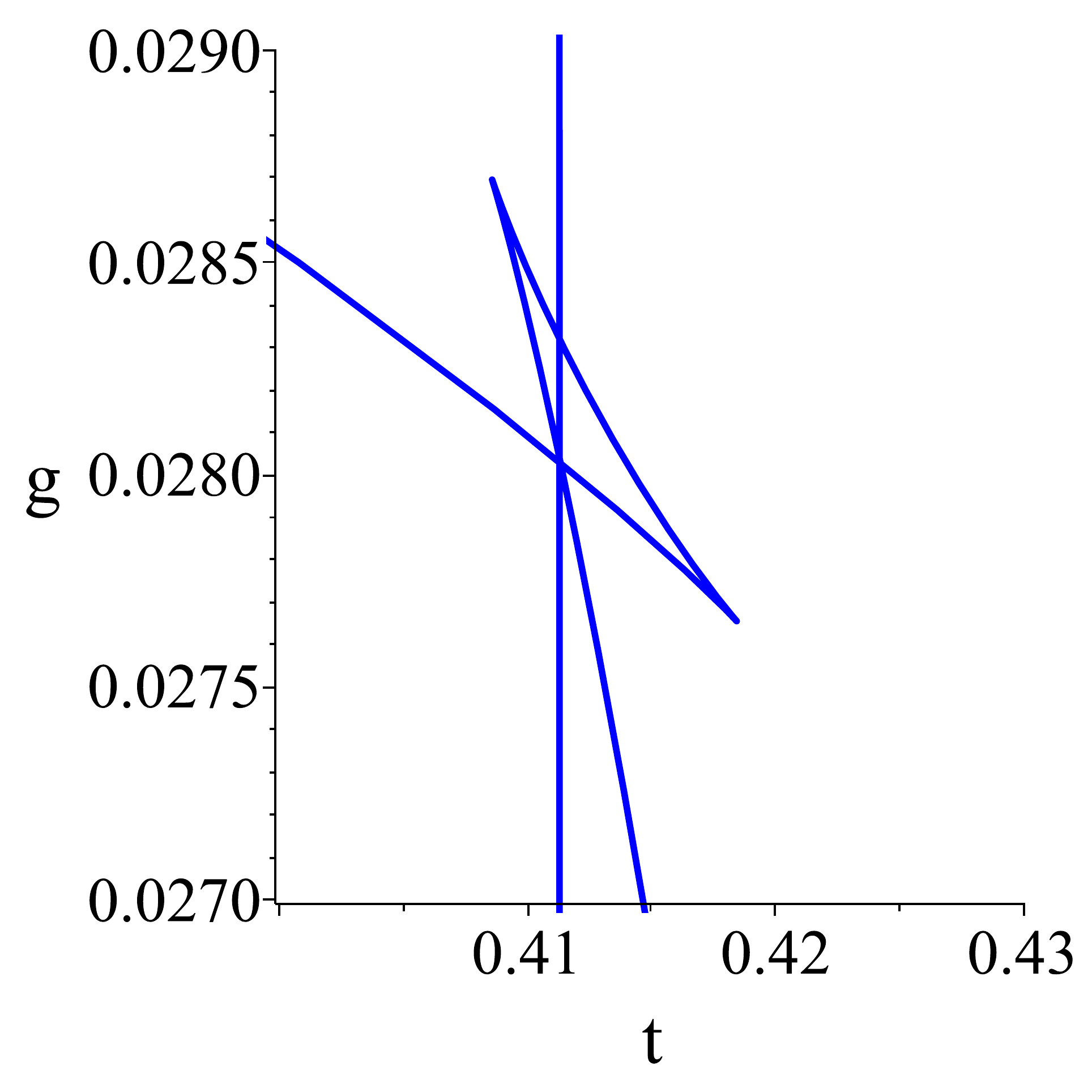}
    \includegraphics[width=0.325\textwidth]{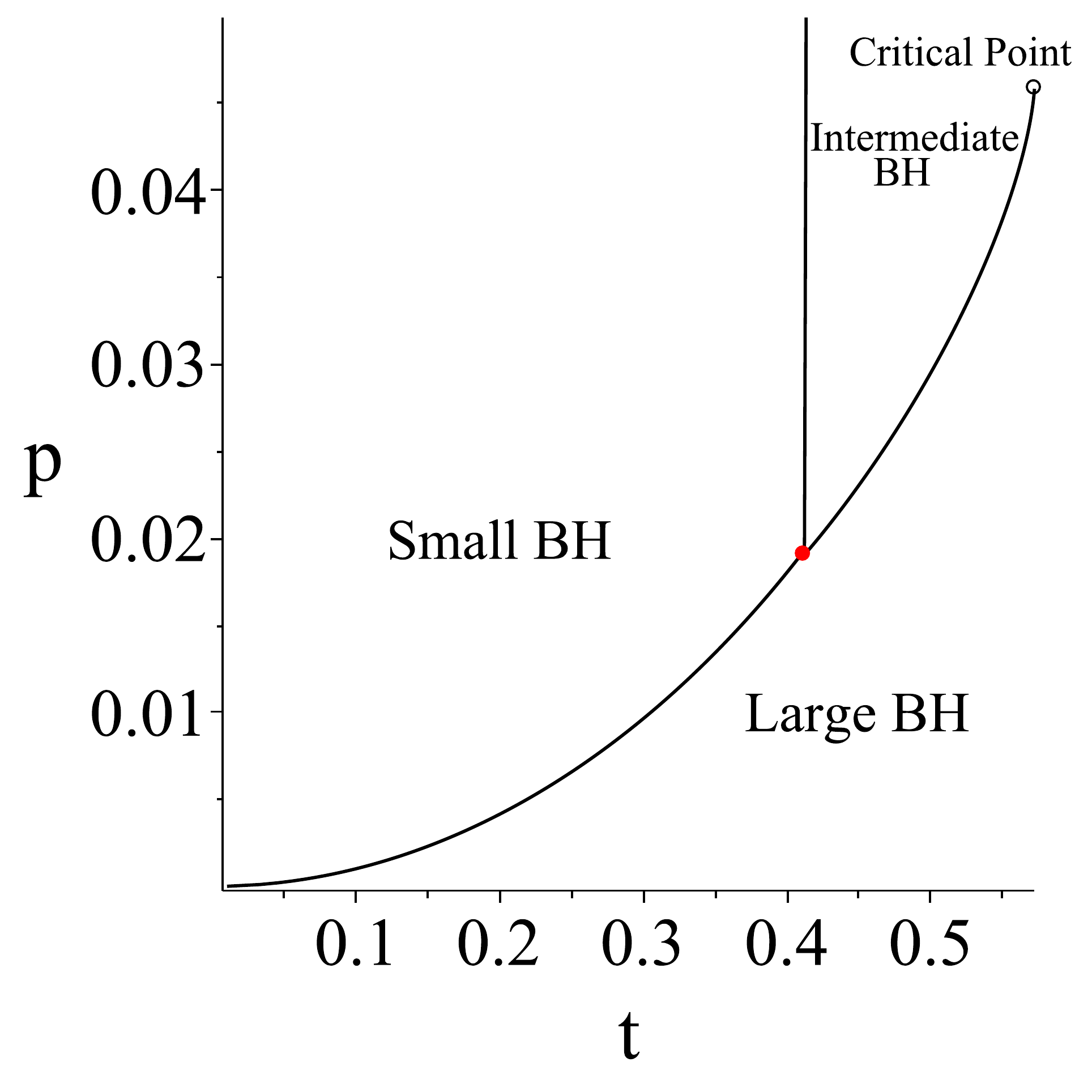}
    \caption{Eight-dimensional charged black hole with $b_{1}=1,\; b_{2}=0.1,\; b_{3}=0.5,\; q=0.05$. \textbf{Left:} $p-v$ diagram for $t\approx0.415$ showing two oscillations. \textbf{Center:} $g-t$ diagram for $p\approx0.19$ showing the intersection of the two swallow tails. \textbf{Right:} Phase diagram showing coexistence lines that all join together at the red triple point. The intermediate-large line will terminate at the critical point. The intermediate/small coexistence line will terminate at the maximum pressure of the spacetime.}
    \label{triple8d}
\end{figure}

Beginning with $b_{2}=0$, when $b_{3} \leq 0$ we observe a single critical point and the corresponding Van der Waals-like transition between a large and small black hole, for all values of $q$. Only quantitative changes in the critical temperature and volume occur when $q$ is varied. When $b_{3}>0$, a triple point may occur for very specific values of $q$ and $b_{3}$. For example, when $b_{3}=0.01$, the standard small-intermediate-large black hole triple point occurs when $0<q\lesssim 0.0069$. When $q\gtrsim 0.0069$ the triple point vanishes and only a small-large transition occurs. On the other hand, when the charge is fixed the existence of the triple point is lower bounded by the parameter $b_3$. For example, when $q=0.05$ we must have $b_{3}>0.4$ for a triple point to occur. This is shown in Figure~\ref{triple8d}.

Next we consider the case where $b_{2}>0$. For any choice of $b_{2}>0$ there is no choice of $q$ and $b_{3}<0$ that yields a triple point. The only phenomena observed are small-large phase transitions. This is also the case when $b_{3}=0$. When $b_{3}>0$, we have the possibility of a small-intermediate-large triple point. As above, the presence of this triple point is highly sensitive to the values of $q$ and $b_{2}$.  If $b_{2}=0.1$ and $b_{3}=0.1$, the triple point only occurs for $0<q<0.006$. If $b_{3}$ is increased while $b_{2}$ is kept fixed, the triple point will occur at a larger value of $q$. Correspondingly, the location of the triple point occurs at a larger temperature $t_c$ and pressure $p_c$. Eventually, if both $b_3$ and $q$ are sufficiently large, the triple point moves above the maximum allowed pressure $p_{max}$, and only a small-large transition is physically allowed. Therefore as long as the maximum pressure bound is not violated, a triple point can occur for any choice of $b_2>0$ and $b_3>0$ provided the black hole has the correct charge.

The behaviour for fixed $b_{3}$ and varying $b_{2}$ and $q$ is qualitatively different. As a concrete example, take $b_{3}=0.1$ and increase $b_{2}$ and $q$ from 0.1 and 0 respectively. A triple point occurs as long as $0.1<b_{2}<0.2$ and $0<q\lesssim0.003$. If {\it both} $b_2\geq0.2$ and $q\gtrsim 0.003$ only a small-large transition occurs. Generically, when the charge {\it decreases} the temperature and pressure at which the triple point occurs will {\it increase}. In all cases, analytic solutions to the locations of the triple/critical points are not available, so the accuracy in the bounds presented are constrained by the choice of numerical solver and precision used.  

The equation of state, free energy, and phase diagram in the presence of a triple point are shown in Figure~\ref{triple8d}. The triple point occurs at the intersection of the small, intermediate, and large black hole branches of the $g-t$ diagram (indicated also by a red dot in the phase diagram). At this point, three distinct black hole states with different sizes all globally minimize the free energy at the same temperature. Figure~\ref{triple8d}c contains the phase diagram, which displays new features compared to the uncharged case. The rightmost branch distinguishes the large and intermediate black hole phases and terminates at a critical point, beyond which the system is in a `superfluid' phase where the two states are no longer distinguishable. The nearly vertical branch separates the intermediate and small black hole phases, and extends upward until it terminates at the maximum allowed pressure $p_{max}$. This is conjectured to always occur before any critical point forms regardless of parameter choice, though a lack of analytic solutions for the location of the critical points prevents us from excluding the possibility of a second critical point occurring below $p_{max}$. Above the critical temperature/pressure which terminates the intermediate-large transition, the two phases are no longer distinguishable (there is no discontinuity in the horizon size). However, we still observe a phase transition between this `merged' intermediate-large black hole branch and the small black hole branch.

\begin{figure}[H]
    \includegraphics[width=0.325\textwidth]{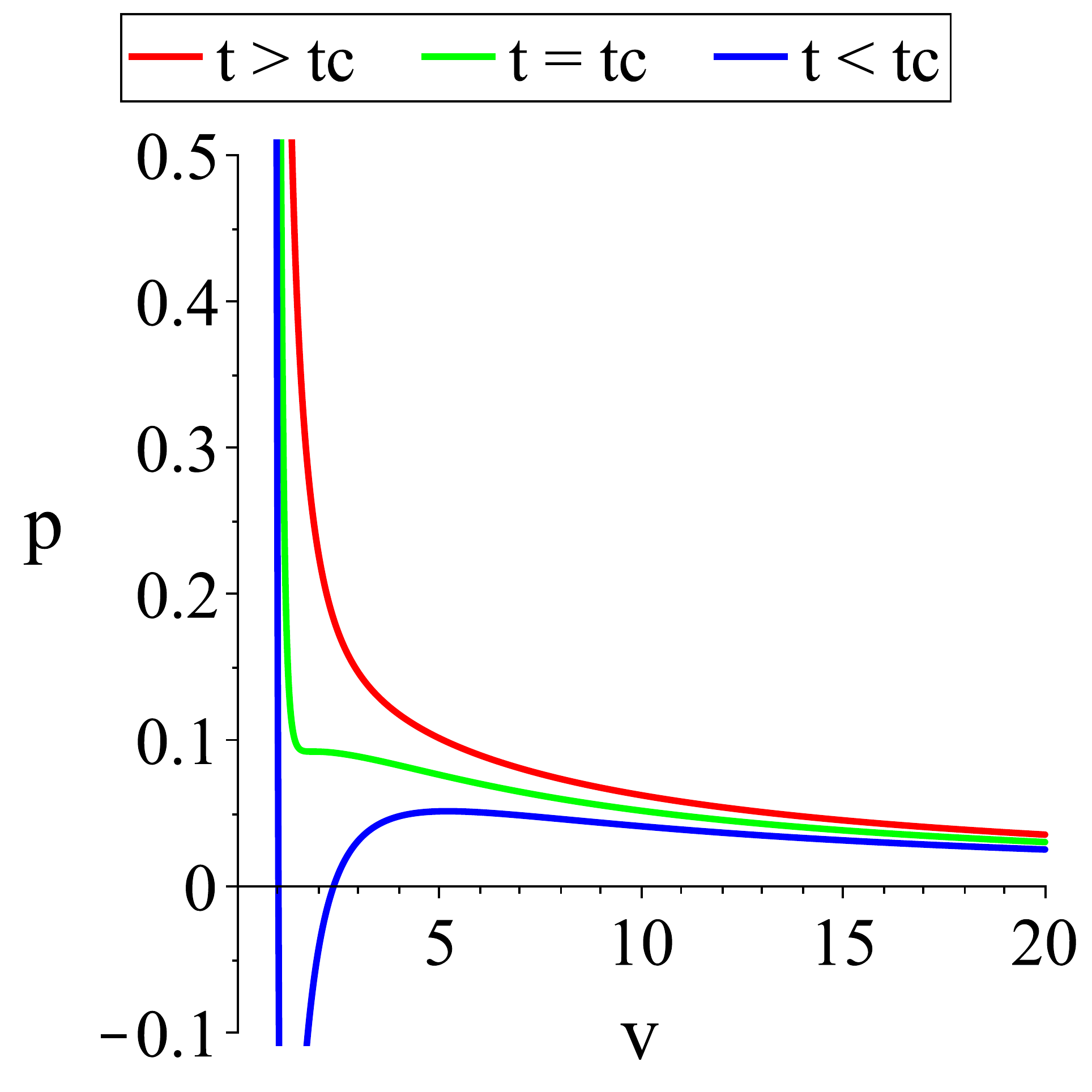}
    \includegraphics[width=0.325\textwidth]{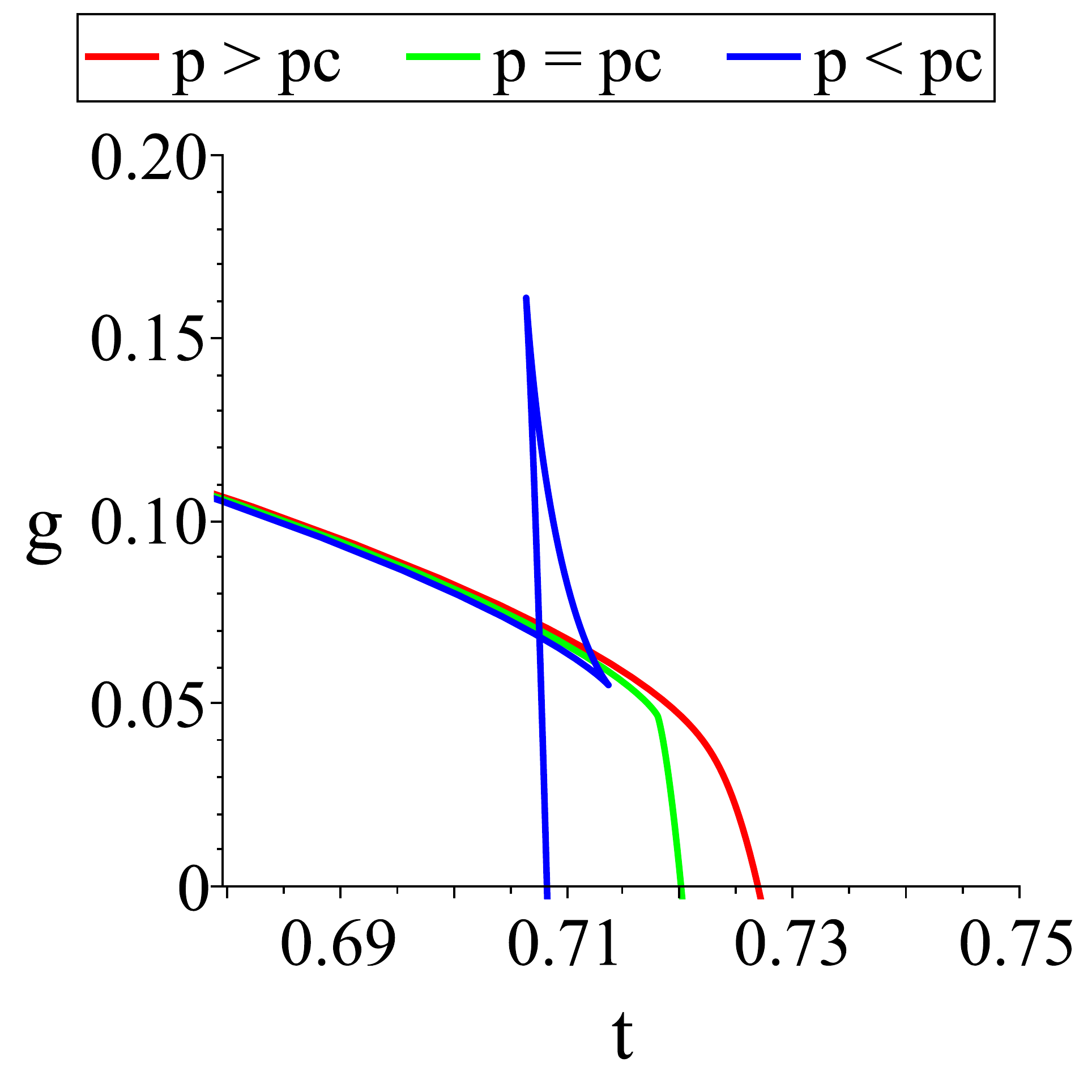}
    \includegraphics[width=0.325\textwidth]{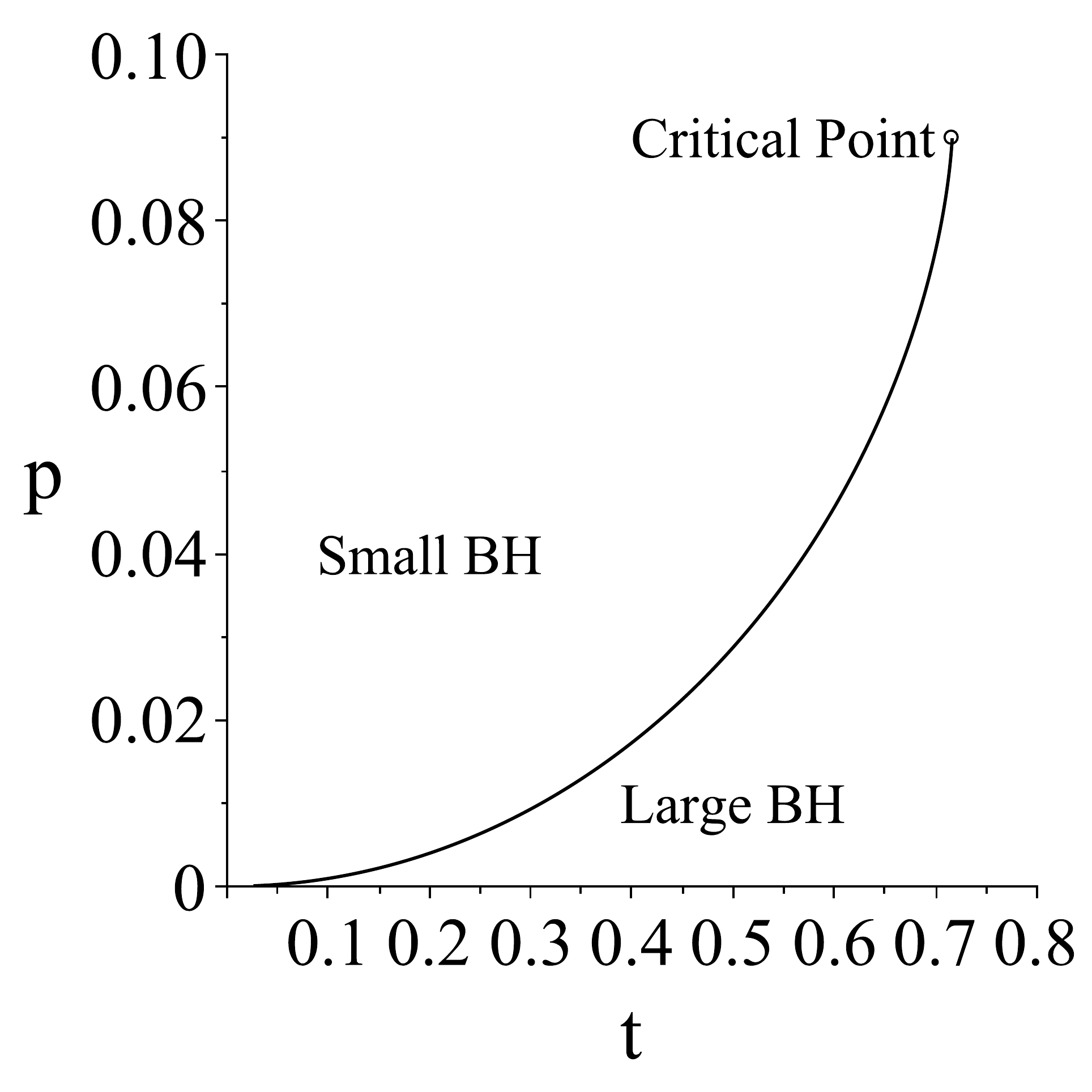}
    \caption{Eight-dimensional charged black holes with $b_{1}=b_{2}=b_{3}=q=1.$ \textbf{Left:} $p-v$ diagram showing oscillation. \textbf{Center:} $g-t$ diagram with standard swallow tail structure. \textbf{Right:} Phase diagram with coexistence line showing the separation of large and small black holes which terminates at the critical point.}
    \label{Sphere8DQ}
\end{figure}

\section{Discussion} \label{conclusion}

Higher curvature (and thus, necessarily higher-dimensional) generalizations of Einstein gravity admit the existence of black hole solutions with `exotic' horizon geometries. In particular, Lovelock gravity allows for black holes which are distinguished by non-constant curvature horizon cross-sections. This is in stark contrast to vacuum solutions in $d=4$ Einstein gravity, which are constrained by Birkhoff's theorem to have transverse metrics isometric to $\mathcal{S}^2$ in the case of spherical symmetry. In this work, we have examined the phase structure of such black hole solutions in third-order Lovelock gravity, focusing on both $U(1)$ charged and uncharged asymptotically AdS solutions in 7 and 8 dimensions and in an extended phase space where the cosmological constant acts as a thermodynamic pressure. Our work can be seen as a natural extension of some preliminary efforts at understanding the thermodynamic properties of these exotic black hole solutions as seen in \cite{farhangkhah2014,hull2021}.

The generalization to third-order Lovelock gravity (and by necessity $d\geq 7$) retains much of the qualitative features of exotic black holes in second-order Gauss-bonnet gravity in $d=5$. Nevertheless we observed and accounted for a wide variety of thermodynamic phenomena which are generic to asymptotically AdS black holes, and found that novel triple point phenomena as seen in \cite{hull2021} persist in higher-order Lovelock gravity. We also clarified the interpretation of earlier results in the case of $d=5$ exotic black holes and $d=7$ constant curvature black holes, and analyzed the space of physically reasonable exotic black hole solutions. 

General `exotic' black hole solutions in third-order Lovelock gravity depend on a set of arbitrary constants  $\{b_{1},b_{2},b_{3}\}$ characterizing the base manifold, as well as the Lovelock coupling constants $\{\alpha_{0},\alpha_{2},\alpha_{3}\}$. We obtained a discriminant condition for the metric function $f(r)$ that establishes a maximum and minimum allowed pressure, which all branches of solution are constrained by. The analysis focused on the Einstein branch of solutions, which has a smooth limit to Einstein gravity when $\alpha_{2},\alpha_{3}\rightarrow 0$. The topological parameters $\{b_n\}$ were treated as constants which can in principle take any values in $\mathbb{R}$. To our knowledge there is currently no known prescription for systematically determining all allowed combinations of topological parameters in third-order Lovelock gravity. Nevertheless, we were able to rule out a large portion of the parameter space by studying how the formation of naked singularities is sensitive to the choice of transverse geometry. We found that most choices of transverse space are pathological, containing a divergence in the Kretschmann scalar at a finite radius outside of the event horizon. This is not surprising as the allowed exotic solutions are already severely constrained in Gauss-Bonnet gravity \cite{Dotti_2005}.

In seven dimensions, we found that uncharged black holes with a \textit{spherical} base manifold contain a novel triple point akin to the one found in \cite{hull2021}, which was not previously seen in the analyses of \cite{frassino2014} and \cite{farhangkhah2014}. This triple point is found to be sensitive to the rescaled Gauss-Bonnet coupling $\alpha$, terminating at the critical point once $\alpha$ reaches a certain value (after which only a Hawking-Page transition occurs). Furthermore, small-large black hole phase transitions are forbidden as they require $\alpha<3$ which precludes the existence of a smooth Einstein limit (see Figure~\ref{fig:alpharegion}). When charge is present, the triple point vanishes and only small-large transitions occur with a single critical point. We clarify that although Hawking-Page transitions were reported previously for this case \cite{farhangkhah2014}, the canonical (fixed-charge) ensemble does not allow for such transitions, even if they appear energetically favourable.

In eight dimensions, black holes without charge exhibit three distinct phase transition behaviours: small-large transitions, Hawking-Page transitions, or triple point phenomena, depending on the choice of topological parameters. For charged black holes, the Hawking-Page transition no longer occurs and we only observe a small-large transition or a triple point. Surprisingly, new types of phase transitions are not seen even with the inclusion of a new topological parameter $b_3$. One might expect a new branch of phase transitions (or even a quadruple point with two intermediate branches) to emerge in $d=8$, because the equation of state contains new terms in higher inverse powers of $r$. It turns out, however, that the constraints present do not allow for choices of parameters which would produce the necessary oscillations/branch structure for such phenomena to occur.

Along the way, we have found solutions for black holes with vanishing mass parameter which nevertheless possess horizons and non-trivial spacetime singularities. These can be found directly from \eqref{ebranch} by appropriate choice of parameters. Investigating the thermodynamic properties and geometric structure of these solutions would be an interesting future endeavour. Another ambitious task would be to examine exotic black holes in asymptotically de-Sitter spaces, where the notion of thermodynamic equilibrium is more subtle. This is an area which has been gaining attention in recent years \cite{sekiwa2006,zhang2014,haroon2020,mbarek2019,simovic2021} with many new thermodynamic features being uncovered for a wide class of black holes. Finally, understanding fully how to constrain the topological parameters would be a significant step towards determining exactly which types of exotic black holes deserve further investigation.

\section*{Acknowledgments}
B.H. is supported in part by the Natural Sciences and Engineering Research Council of Canada. F.S is supported by the Australian Research Council. We are grateful to Robert B. Mann for a number of helpful discussions.

\newpage
\bibliographystyle{ieeetr}
\bibliography{LBIB.bib}

\end{document}